\documentclass[pdflatex,sn-mathphys-num]{sn-jnl}
\usepackage{float}
\usepackage{amsmath,amssymb,amsfonts}%
\usepackage{amsthm}%
\usepackage{float}
\usepackage{textcomp}
\usepackage{float}
\usepackage{graphicx} 
\usepackage{comment}
\usepackage{amsmath}
\usepackage{amssymb}
\usepackage{threeparttable}
\usepackage{placeins}
\usepackage{subcaption}
\usepackage{hyperref} 
\usepackage{makecell}  

\title{Exploring Emergent Topological Properties in Socio-Economic Networks through Learning Heterogeneity}
\author{}

\begin{document}

\abstract{Understanding how individual learning behavior and structural dynamics interact is essential to modeling emergent phenomena in socio-economic networks. While bounded rationality and network adaptation have been widely studied, the role of heterogeneous learning rates—both at the agent and network levels—remains underexplored. This paper introduces a dual-learning framework that integrates individualized learning rates for agents and a rewiring rate for the network, reflecting real-world cognitive diversity and structural adaptability.

Using a simulation model based on the Prisoner's Dilemma and Quantal Response Equilibrium, we analyze how variations in these learning rates affect the emergence of large-scale network structures. Results show that lower and more homogeneously distributed learning rates promote scale-free networks, while higher or more heterogeneously distributed learning rates lead to the emergence of core-periphery topologies. Key topological metrics—including scale-free exponents, Estrada heterogeneity, and assortativity—reveal that both the speed and variability of learning critically shape system rationality and network architecture. This work provides a unified framework for examining how individual learnability and structural adaptability drive the formation of socio-economic networks with diverse topologies, offering new insights into adaptive behavior, systemic organization, and resilience.}


\author[1]{\fnm{Chanuka} \sur{Karavita}}\email{chanuka.k@sliit.lk}

\author[2]{\fnm{Zehua} \sur{Lyu}}\email{zlyu0064@uni.sydney.edu.au}

\author*[1]{\fnm{Dharshana} \sur{Kasthurirathna}}\email{dharshana.k@sliit.lk}

\author*[2]{\fnm{Mahendra} \sur{Piraveenan}}\email{mahendrarajah.piraveenan@sydney.edu.au}

\affil[1]{\orgdiv{Faculty of Computing}, \orgname{Sri Lanka Institute of Information Technology}, \orgaddress{\city{Malabe}, \country{Sri Lanka}}}

\affil[2]{\orgdiv{School of Computer Science, Faculty of Engineering}, \orgname{The University of Sydney}, \orgaddress{\city{Sydney, NSW}, \country{Australia}}}

\maketitle
 

\newpage

\section{Introduction}

The study of emergent behavior in complex systems has garnered substantial attention across disciplines ranging from physics and computer science to sociology and economics. Network science, in particular, has emerged as a foundational paradigm for understanding how large-scale structural patterns—such as scale-free degree distributions, modularity, and star-like or centralized topologies—arise from local interactions among individual agents \cite{barabasi1999emergence, newman2010networks}. Within socio-economic systems, these emergent structures are not merely artifacts of connectivity, but are intricately shaped by the behavioral rules governing agent decision-making. As such, modeling the interplay between individual cognition and network evolution is critical for capturing the adaptive dynamics of real-world systems \cite{jackson2010social, castellano2009statistical}.

Bounded rationality, originally introduced by Herbert Simon \cite{simon1955behavioral}, offers a more realistic alternative to the idealized assumptions of perfect rationality in classical game theory. In bounded rationality models, agents operate under cognitive and informational constraints, and their decision-making is often probabilistic and adaptive rather than deterministic and optimal. When agents embedded in a network play strategic games—such as the Prisoner's Dilemma—their local updates, combined with evolving connectivity, can give rise to global patterns of cooperation, centralization, or fragmentation \cite{nowak1992evolutionary, perc2017statistical}. This has been shown to drive important phenomena in domains such as opinion formation, financial stability, and collective decision-making \cite{acemoglu2010spread, lamba2021emergence, piraveenan2021optimal}.

Despite the growing interest in bounded rationality and evolutionary game dynamics on networks, most existing models assume uniform cognitive adaptability across agents and static or uniformly updated network structures \cite{galla2013complex, candogan2011fluctuations}. However, real-world populations are rarely homogeneous. Individuals differ in how quickly they learn, how sensitive they are to incentives, and how much weight they give to feedback—due to psychological, socio-economic, or cultural factors \cite{hofstede2001culture, cowan2014working, roberts2006personality}. Likewise, some systems (e.g., bureaucracies or centralized platforms) adapt structurally much faster or slower than others (e.g., grassroots communities) \cite{rogers2003diffusion, zahra2002absorptive}. Despite being well-documented in cognitive psychology, organizational theory, and cultural studies, these forms of heterogeneity—such as individual differences in learnability and systemic variation in structural adaptability—remain largely absent from models of bounded rationality in dynamic, networked game-theoretic frameworks.

This paper addresses this gap by introducing a novel dual-learning framework that captures the heterogeneity of learnability at both the node and network levels. At the node level, we assign individualized learning rates to agents, representing the rate at which they adapt their rationality based on local feedback. At the network level, we model rewiring rate—how often the network structure adapts in response to systemic performance—as a form of collective learning. Through a series of simulation experiments, we demonstrate that this joint heterogeneity gives rise to diverse emergent topologies, including scale-free and core-periphery structures. These outcomes closely mirror real-world networks such as global airline systems, financial interbank markets, and urban transportation infrastructures \cite{guimera2005worldwide, boss2004core}, where both individual learnability and structural adaptability are highly variable in nature \cite{piraveenan2020topology}.

\subsection*{Contributions}

This work makes several key contributions:
\begin{itemize}
    \item It introduces a dual-axis learning framework that models agent rationality updates and network rewiring as distinct but interacting learning processes.
    \item It incorporates heterogeneous learning rates for agents, drawn from normal distributions with varying standard deviations, to reflect cognitive diversity.
    \item It operationalizes network-level adaptation through a rewiring rate parameter \( F \), capturing how frequently the topology is re-evaluated.
    \item It provides empirical insights into how learning heterogeneity influences the emergence of socio-economic systems with varying topologies, using metrics such as scale-free exponents, Estrada heterogeneity, and assortativity.
\end{itemize}

\subsection*{Structure of the paper}
The remainder of the paper is organized as follows. Section~\ref{background} reviews relevant literature on bounded rationality, Quantal Response Equilibrium, and heterogeneous learning in socio-economic systems. Section~\ref{sec:methodology} details the simulation framework, including the learning and rewiring models, experimental configurations, and evaluation metrics. Section~\ref{sec:results} presents the results of simulations under varying learning and rewiring rates, with a focus on how heterogeneity influences emergent topologies and rationality. Section~\ref{sec:discussion} interprets the findings in light of real-world systems, outlines theoretical implications, and discusses limitations and directions for future research. Finally, the paper concludes by highlighting the broader significance of modeling learning heterogeneity in adaptive networks.

\section{Background}
\label{background}
The study of complex networks has provided key insights into how large-scale structures and behaviors emerge from local interactions among individual agents. These emergent patterns---ranging from scale-free degree distributions to core-periphery topologies---are particularly important in socio-economic systems, where agents continuously adapt to their environment while interacting over dynamic network structures \cite{barabasi1999emergence, watts1998collective, boccaletti2006complex}.

\subsection{Bounded Rationality and Emergent Behavior in Networked Games}

Bounded rationality, a concept introduced by Herbert Simon \cite{simon1955behavioral}, describes how individuals make decisions under cognitive and informational constraints. In contrast to the fully rational agents of classical economic theory, boundedly rational agents rely on heuristics, have limited foresight, and update their strategies gradually based on feedback. When such agents are embedded in networks and play repeated strategic games, their interactions can drive the formation of large-scale patterns such as modularity, clustering, and long-tailed connectivity \cite{jackson2010social, gross2008adaptive}.

To study this, researchers have employed networked versions of classic games such as the Prisoner's Dilemma (PD). In our formulation, each agent plays a PD game with its neighbors using the following payoff matrix, where $\delta > \alpha > \beta > \gamma$:

\begin{center}
\begin{tabular}{|c|c|c|}
\hline
 & Cooperate & Defect \\
\hline
Cooperate & $(\alpha, \alpha)$ & $(\gamma, \delta)$ \\
\hline
Defect & $(\delta, \gamma)$ & $(\beta, \beta)$ \\ 
\hline
\end{tabular}
\end{center}
 
This setup ensures that defection is individually optimal but mutual cooperation would yield a better collective outcome. These tensions between local incentives and global performance are central to the emergence of network-level rationality \cite{szabo2007evolutionary, perc2010coevolutionary}.

\subsection{Quantal Response Equilibrium and System Rationality}

To account for probabilistic bounded decision-making, the Quantal Response Equilibrium (QRE) model, developed by McKelvey and Palfrey \cite{mckelvey1995quantal}, is widely adopted. In QRE, agents do not always choose the best strategy deterministically. Instead, they choose actions with a probability that increases with expected utility, modulated by a rationality parameter $\lambda_i$:

\begin{equation}
P_i(a) = \frac{e^{\lambda_i U_i(a)}}{\sum_b e^{\lambda_i U_i(b)}}
\end{equation}

Here, $U_i(a)$ is the expected utility of action $a$, and $\lambda_i$ captures how "rational" agent $i$ is. As $\lambda_i \to \infty$, the agent behaves perfectly rationally. When $\lambda_i = 0$, the agent acts completely at random. This parameter is thus central to modeling bounded rationality in both prior work and the current study \cite{goeree2001ten, goeree2016quantal}.

System rationality is defined as the network-wide average closeness of agent behavior to Nash Equilibrium. Following Kasthurirathna et al. \cite{kasthurirathna2015emergence}, system rationality $\rho(\mathcal{N})$ is calculated as the negative average Jensen-Shannon (JS) divergence between each agent pair's QRE strategy distributions $P_k$ and corresponding Nash strategies $Q_k$:

\begin{equation}
\rho(\mathcal{N}) = -\frac{1}{M} \sum_{k=1}^{M} \text{JS}(P_k || Q_k)
\end{equation}

where $M$ is the number of edges. The JS divergence is computed as:

\begin{equation}
\text{JS}(P || Q) = \frac{1}{2} \sum_i P_i \log \frac{P_i}{M_i} + \frac{1}{2} \sum_i Q_i \log \frac{Q_i}{M_i}
\end{equation}

with $M_i = \frac{1}{2}(P_i + Q_i)$. A higher $\rho(\mathcal{N})$ (i.e., lower JS divergence) indicates a more rational, coherent system.

\subsection{Learning Rates in Adaptive Systems}

In adaptive systems, learning rates determine how fast agents revise internal parameters based on feedback. A high learning rate results in rapid but potentially unstable adaptation, whereas a low learning rate allows for slower, more stable learning.

In gradient descent \cite{bottou1998online}, the parameter update rule is:

\begin{equation}
\theta_{t+1} = \theta_t - \eta \nabla L(\theta_t)
\end{equation}

where $\eta$ is the learning rate. In reinforcement learning, the Bellman update \cite{sutton2018reinforcement} with learning rate $\alpha$ is given by:

\begin{equation}
V_{t+1}(s) = (1 - \alpha)V_t(s) + \alpha [R(s) + \gamma \max_a V(s')]
\end{equation}

This structure is mirrored in our model where agent rationality evolves as a function of an individual learning rate and an expected rationality target, as detailed in the Methodology section.

\subsection{Heterogeneity in Learning Rates Across Individuals and Societies}

Empirical research across disciplines supports the existence of heterogeneous learning rates. In psychology, factors such as age, working memory, and personality impact learnability \cite{cowan2014working, roberts2006personality}. Hofstede's cultural dimensions \cite{hofstede2001culture} suggest that national differences in uncertainty avoidance and long-term orientation influence how populations respond to change.


Recent empirical and simulation-based analyses have highlighted how variation in learning speeds among individuals can substantially influence cooperation dynamics and the structural evolution of socio-economic networks \cite{rahwan2019machine}. Evidence also suggests that such heterogeneity in learning can affect equilibrium selection and shape the emergence of specific network patterns \cite{merhej2021cooperation}. These observations support the integration of agent-level differences in rational adaptation within modeling frameworks. Moreover, the conceptual foundation of this work aligns with theoretical developments in organizational learning, where learning is viewed as a multi-level process involving both behavioral change and structural reconfiguration over time \cite{miner2020organizational}. This perspective reinforces the broader relevance of the proposed dual-learning simulation framework for capturing co-evolutionary processes in complex socio-economic systems.

In organizational and innovation studies, differences in absorptive capacity \cite{zahra2002absorptive} and diffusion categories \cite{rogers2003diffusion} highlight structural variation in learning responsiveness. These findings justify modeling both node-level learning rates $\alpha_i$ and a network-level learning rate $F$ (rewiring rate), as proposed in this work.

\subsection{Metrics for Topological Analysis}

To analyze network evolution, we use the following topological metrics:

\begin{itemize}
  \item \textbf{Scale-Free Correlation:} Goodness-of-fit between empirical and theoretical power-law distributions \cite{clauset2009power}.
  \item \textbf{Scale-Free Exponent ($\gamma$):} Slope of degree distribution tail; $2 < \gamma < 3$ indicates scale-free behavior \cite{barabasi1999emergence}.
  \item \textbf{Estrada Heterogeneity Index} \cite{estrada2010quantifying}:
  \begin{equation}
  H_E = \frac{1}{n} \sum_{i=1}^n \left( \frac{e_i}{\sum_j e_j} \right)^2
  \end{equation}
  where $e_i$ is node communicability or eigenvalue centrality. It helps capture the emergence of inequality in node roles and influence, reflecting centralization.
  \item \textbf{Degree Assortativity:} Pearson correlation between degrees of linked nodes \cite{newman2002assortative}.  It measures the tendency of nodes to connect to other nodes with similar degree \cite{piraveenan2010assortative, piraveenan2008local}.
  \item \textbf{Rationality Assortativity:} Correlation between node rationality $\lambda_i$ and that of connected neighbors. A drop in rationality assortativity means rational agents are becoming hubs, guiding or dominating less rational ones.
\end{itemize}

These metrics not only characterize network properties individually, but together they offer strong empirical indicators of underlying network structure. For example, low scale-free exponents (e.g., $\gamma \lesssim 1.6$) indicate heavy-tailed, hub-dominated networks that depart from classical preferential attachment processes \cite{barabasi1999emergence}. Estrada’s heterogeneity index captures skewed communicability and centrality concentration, which are typical of core-periphery structures \cite{estrada2010quantifying}. Likewise, disassortative mixing—as revealed by negative degree or rationality assortativity—suggests a structural split between well-connected cores and loosely connected peripheries \cite{borgatti2000models, newman2002assortative}.

\section{Methodology}
\label{sec:methodology}

The simulation framework employed in this study is based on a simulated annealing approach, inspired by prior work on adaptive socio-economic networks by Kasthurirathna and Piraveenan \cite{kasthurirathna2015emergence}, as well as similar models in the literature \cite{szabo2007evolutionary, roman2017topology}. These models emphasize co-evolutionary dynamics, where both agent strategies and network topologies adapt in response to performance signals, enabling the emergence of complex collective behavior from simple local rules \cite{kasthurirathna2015influence, kasthurirathna2016modeling, kasthurirathna2016optimising, kasthurirathna2022disassortative}.

Our work specifically extends these approaches by integrating two distinct learning dynamics: (1) \textbf{node-level learning rates}, which govern the pace at which individual agents adapt their behavior, and (2) \textbf{network-level learning rates}, represented by a rewiring rate parameter $F$, which determines how frequently the structure of the network is updated in response to system rationality feedback.

\subsection{Simulation Framework}

We simulate an adaptive network consisting of 1000 nodes, initialized as a connected Erd\H{o}s--R\'enyi random graph with 3000 edges \cite{erdos1959random}. Agents interact over these edges via a pairwise Prisoner's Dilemma game, using the following payoff matrix: $\delta = 5$, $\alpha = 3$, $\beta = 1$, and $\gamma = 0$, where $\delta > \alpha > \beta > \gamma$ ensures that defection is the dominant strategy in a one-shot interaction \cite{rapoport1965prisoner}.

Each agent's strategy follows the Quantal Response Equilibrium (QRE) framework, as defined in Section~\ref{background}, using agent-specific rationality $\lambda_i$ \cite{mckelvey1995quantal}.

\subsection{Node-Level Learning Rate and Rationality Update}

Each agent possesses a personalized \textbf{node-level learning rate} $\alpha_i$, which governs how rapidly it updates its rationality. The learning rate is a scalar in the range $[0, 1]$, where $\alpha_i = 1$ indicates immediate convergence to the target rationality (perfect learning), and $\alpha_i = 0$ implies no learning. This parameter captures the heterogeneity in cognitive adaptability across a population \cite{cowan2014working}.

Rationality evolves over time according to a recursive update:
\begin{equation}
\lambda_i(t+1) = (1 - \alpha_i)\lambda_i(t) + \alpha_i \lambda_i^{\text{exp}}(t)
\end{equation}



The expected rationality $\lambda_i^{\text{exp}}(t)$ is defined using two alternative functions of the node degree $d_i(t)$ to improve the generalizability of results across different structural sensitivity assumptions. The two formulations are as follows:

\begin{equation}
\lambda_i^{\text{exp}}(t) =
\begin{cases}
e^{d_i(t)^3}, & \text{(exponential-cubic form)} \\
d_i(t)^2,     & \text{(polynomial-quadratic form)}
\end{cases}
\end{equation}

The exponential-cubic function strongly amplifies rationality differences between low and high-degree nodes, making it highly sensitive to structural position. In contrast, the polynomial-quadratic form offers a milder, yet still nonlinear, scaling of rationality based on degree. These two alternatives allow for broader exploration of how node connectivity influences rational behavior. While other node-level metrics—such as clustering coefficient, eigenvector centrality, or betweenness centrality—could also serve as inputs to $\lambda_i^{\text{exp}}(t)$ \cite{freeman1977set}, degree was chosen for its computational efficiency and clear interpretability across network topologies.

 
\subsection{Network-Level Learning via Rewiring and Simulated Annealing}

The network structure is dynamically updated through a simulated annealing-based rewiring mechanism \cite{kirkpatrick1983optimization}. At periodic intervals, a subset of links is considered for rewiring to evaluate whether an alternative configuration would yield higher system rationality, as defined in Section~\ref{background}. If the proposed rewiring improves global rationality, it is accepted. Otherwise, it may still be accepted with a certain probability based on a decaying temperature function $T(t)$, following the principles of simulated annealing.

The \textbf{rewiring rate} $F \in (0,1]$ serves as the \textbf{network-level learning rate}, determining how often the network checks for and executes potential rewiring actions. Formally, if $F = 0.1$, it implies that the structure is evaluated for rewiring every $\frac{1}{F} = 10$ iterations. Thus, lower values of $F$ correspond to slower structural adaptation (less frequent rewiring), whereas higher values of $F$ result in faster topological learning. This parameter complements the individual-level learning rates $\alpha_i$, capturing the two axes of co-evolution in the system: cognitive adaptation and structural plasticity.

\newpage
\subsection{Experimental Configurations}

We examine three primary experiment families:

\textbf{1. Uniform Learning Rates:} All agents share the same learning rate $\alpha$. Multiple trials are run with $\alpha = 0.1, 0.3, 0.5, 0.7, 0.9$ to analyze the role of adaptation speed.

\textbf{2. Normally Distributed Learning Rates:} In this configuration, each agent's learning rate $\alpha_i$ is sampled from a normal distribution $\mathcal{N}(\mu, \sigma^2)$, where $\mu$ is the average learning rate and $\sigma$ (the standard deviation) controls the spread or variability of learning rates across the population. A smaller standard deviation (e.g., $\sigma = 0.01$) results in a population with nearly uniform learning capacities, while a larger standard deviation (e.g., $\sigma = 0.1$) introduces significant heterogeneity---allowing some agents to adapt much faster or slower than others. This setup is used to explore how intra-population variability in adaptability shapes global network outcomes \cite{rogers2003diffusion}.

\textbf{3. Rewiring Rate Experiments:} We evaluate multiple values of the rewiring rate $F = 0.1, 0.2, 0.5, 1.0$ to examine how the frequency of structural adaptation affects network evolution. These trials isolate the influence of the network-level learning process \cite{gross2008adaptive}. The exponential learning rate function was employed in all the experiments. 

All experimental results were averaged over $10$ independent simulation runs to ensure robustness and reduce the impact of stochastic variability.

\subsection{Evaluation Metrics}

We use several key metrics to analyze results:
\begin{itemize}
    \item \textbf{Average Node Rationality:} Mean $\lambda_i$ across all nodes.
    \item \textbf{System Rationality:} Computed using the Jensen-Shannon divergence between QRE and Nash strategies, as described in Section~\ref{background} \cite{kasthurirathna2015emergence}.
    \item \textbf{Scale-Free Correlation:} Indicates goodness-of-fit to a power-law distribution \cite{clauset2009power}.
    \item \textbf{Scale-Free Exponent} ($\gamma$): Slope of the degree distribution tail, typically between 2 and 3 for scale-free behavior \cite{barabasi1999emergence}.
    \item \textbf{Estrada Heterogeneity Index:} Captures inequality in node communicability, based on spectral characteristics of the network \cite{estrada2010quantifying}.
    \item \textbf{Rationality and Degree Assortativity:} Pearson correlation between connected nodes' rationalities and degrees, respectively, to assess homophily or disassortativity trends \cite{newman2002assortative}.
\end{itemize}

All the simulations were run over one thousand iterations and the intermediate networks were used to quantify and observe the evolving topological properties. The source code used to implement the simulation and analysis is publicly available at the following GitHub repository: \href{https://bit.ly/4lIFCBi}{https://bit.ly/4lIFCBi}

\newpage
\section{Results}
\label{sec:results}

This section presents the results of simulation experiments designed to explore how varying learning rates and rewiring rates influence network evolution and system rationality. The results show that both the speed and variability of learning, as well as the frequency with which the network rewires, have substantial effects on the resulting topology.

In all configurations, networks tend to evolve into either \textit{core-periphery} or \textit{scale-free} structures. The distinction between these is primarily based on:
\begin{itemize}
    \item The \textbf{scale-free exponent} ($\gamma$) of the degree distribution.
    \item The \textbf{Estrada heterogeneity index}, which quantifies inequality in centralities of the nodes.
\end{itemize}

\FloatBarrier

Although the scale-free correlation confirms whether the degree distribution follows a power law (i.e., whether the network is scale-free at all), it does not differentiate between shallow vs. steep distributions, a key distinction between organic scale-free and hub-dominated topologies.

\subsection{Topological Changes under uniform Learning Rates}
 

   \begin{figure*}[htbp]
        \centering
        \begin{subfigure}[t]{0.45\textwidth}
            \centering
            \includegraphics[width=\linewidth]{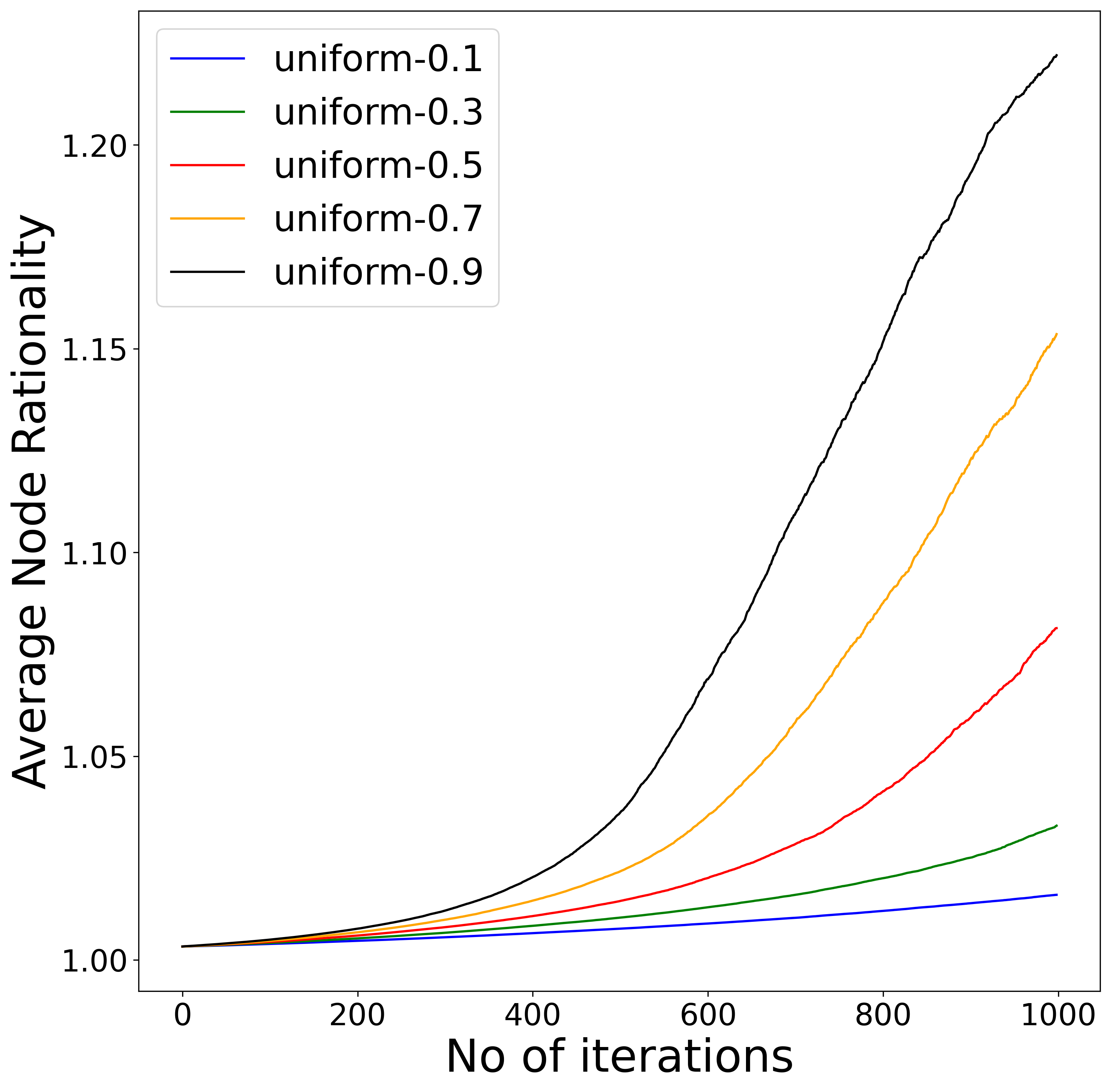}
            \caption{Exponential}
            \label{uniform_avgR_1}
        \end{subfigure}
        \hfill
        \begin{subfigure}[t]{0.45\textwidth}
            \centering
            \includegraphics[width=\linewidth]{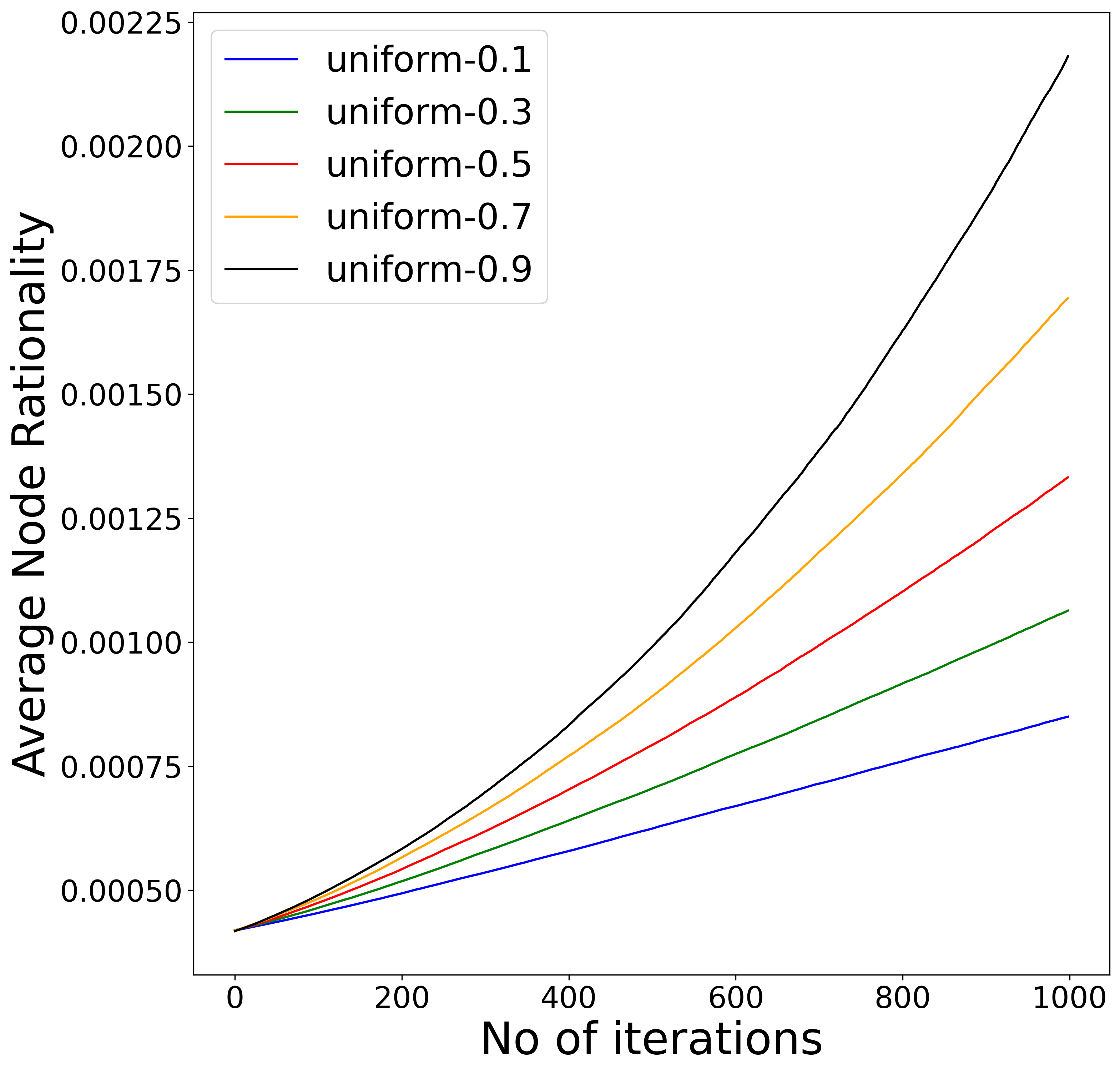}
            \caption{Polynomial}
            \label{uniform_avgR_2}
        \end{subfigure}
        \caption{Average node rationality under uniform learning rates.}
        \label{uniform_avgR}
    \end{figure*}

\FloatBarrier
\newpage
\begin{figure*}[htbp]
    \centering
    \begin{subfigure}[t]{0.45\textwidth}
        \centering
        \includegraphics[width=\linewidth]{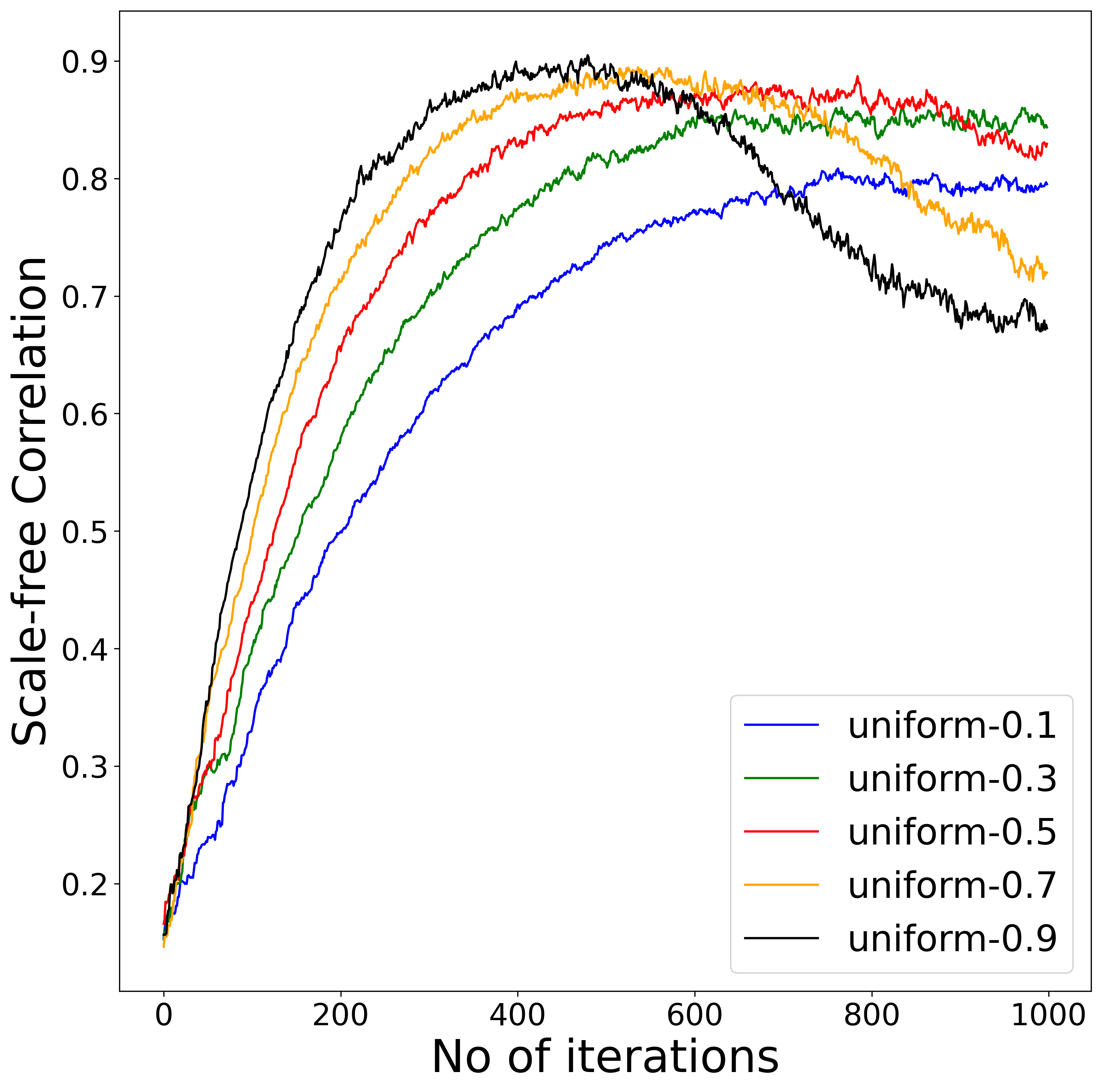}
        \caption{Exponential}
        \label{uniform_Scalefree_1}
    \end{subfigure}
    \hfill
    \begin{subfigure}[t]{0.45\textwidth}
        \centering
        \includegraphics[width=\linewidth]{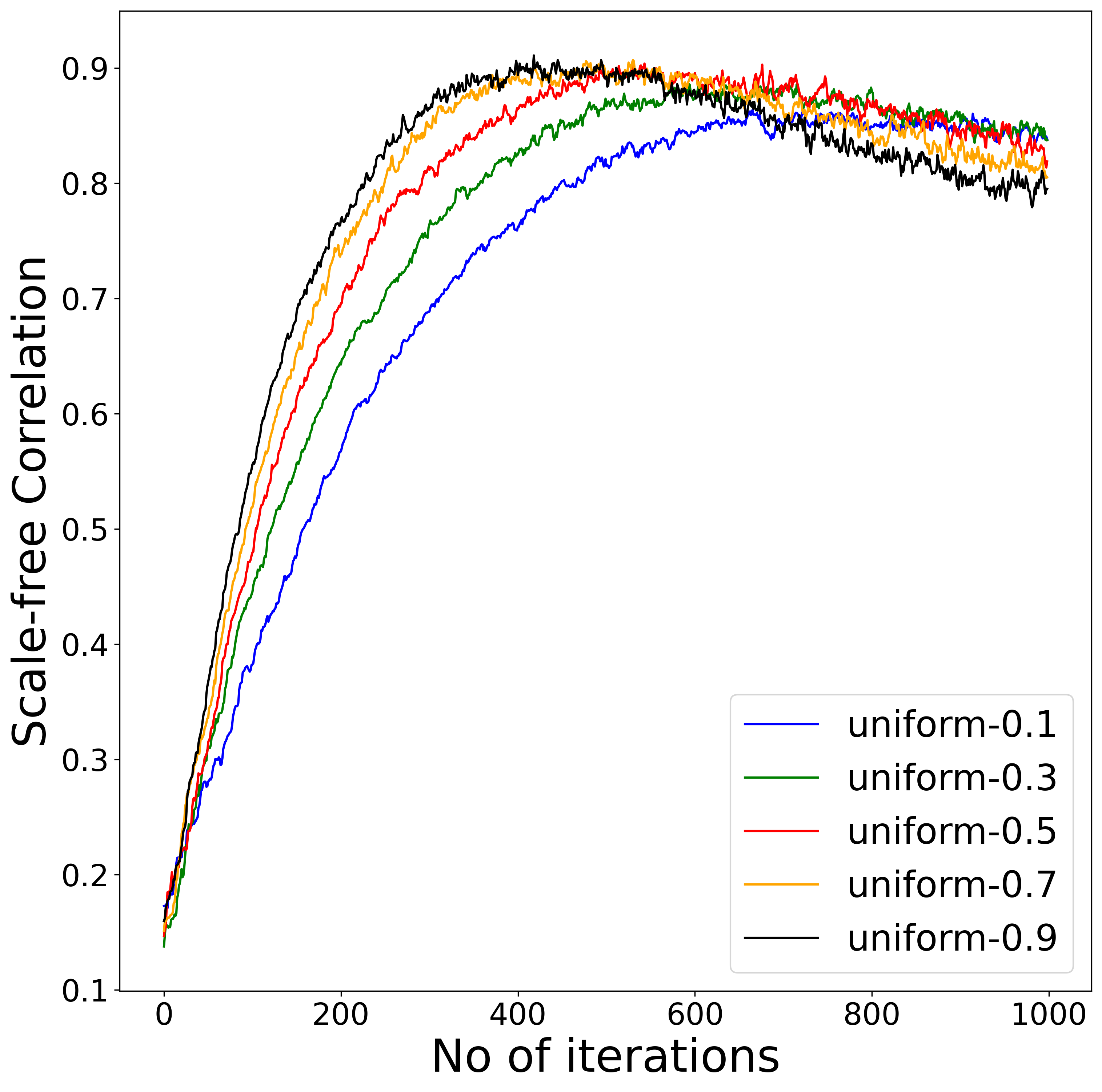}
        \caption{Polynomial}
        \label{uniform_Scalefree_2}
    \end{subfigure}
    \caption{Scale-free correlation under uniform learning rates.}
    \label{uniform_Scalefree}
\end{figure*}

\FloatBarrier

\begin{figure*}[htbp]
    \centering
    \begin{subfigure}[t]{0.45\textwidth}
        \centering
        \includegraphics[width=\linewidth]{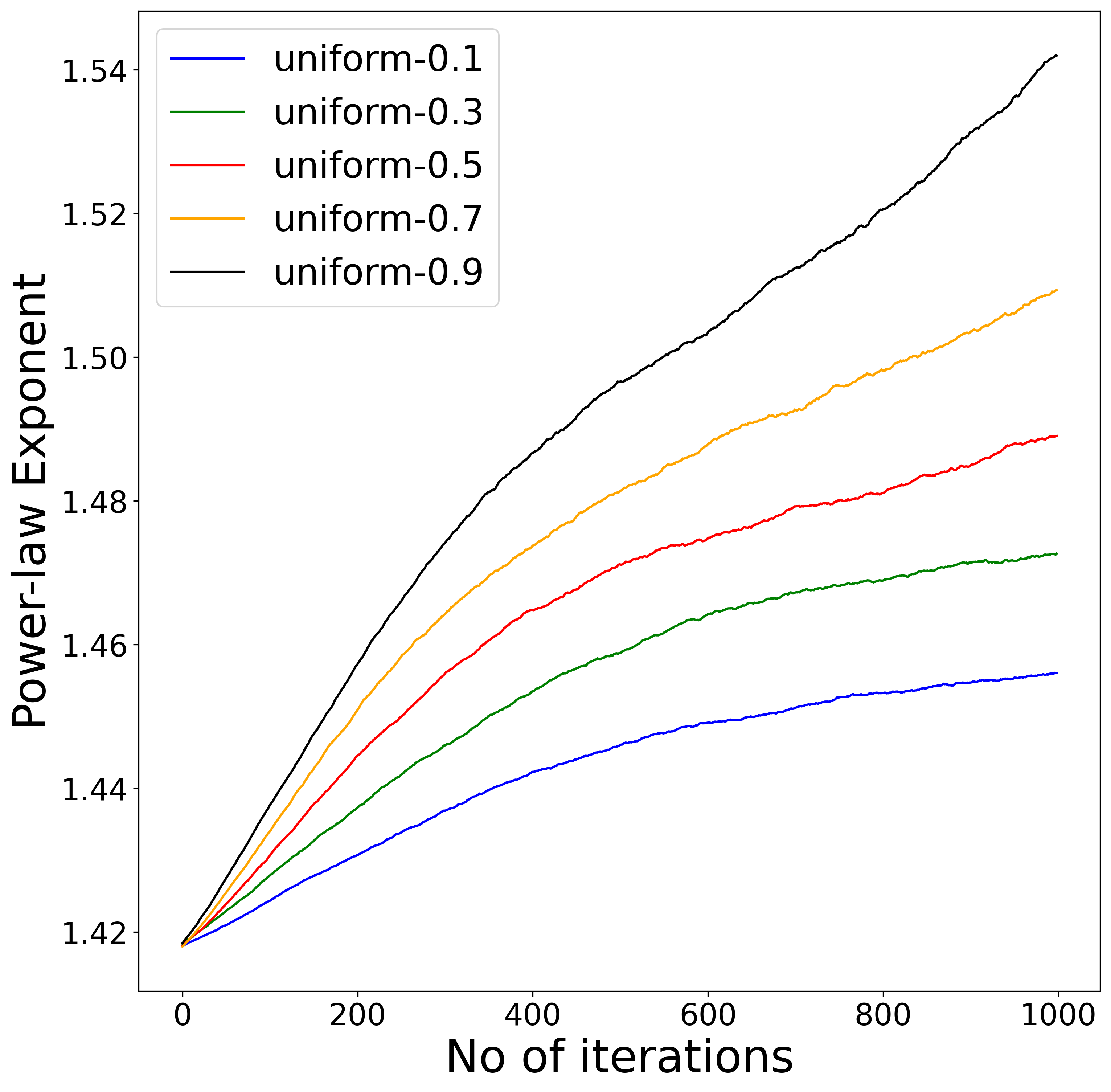}
        \caption{Exponential}
        \label{uniform_Powerlaw_1}
    \end{subfigure}
    \hfill
    \begin{subfigure}[t]{0.45\textwidth}
        \centering
        \includegraphics[width=\linewidth]{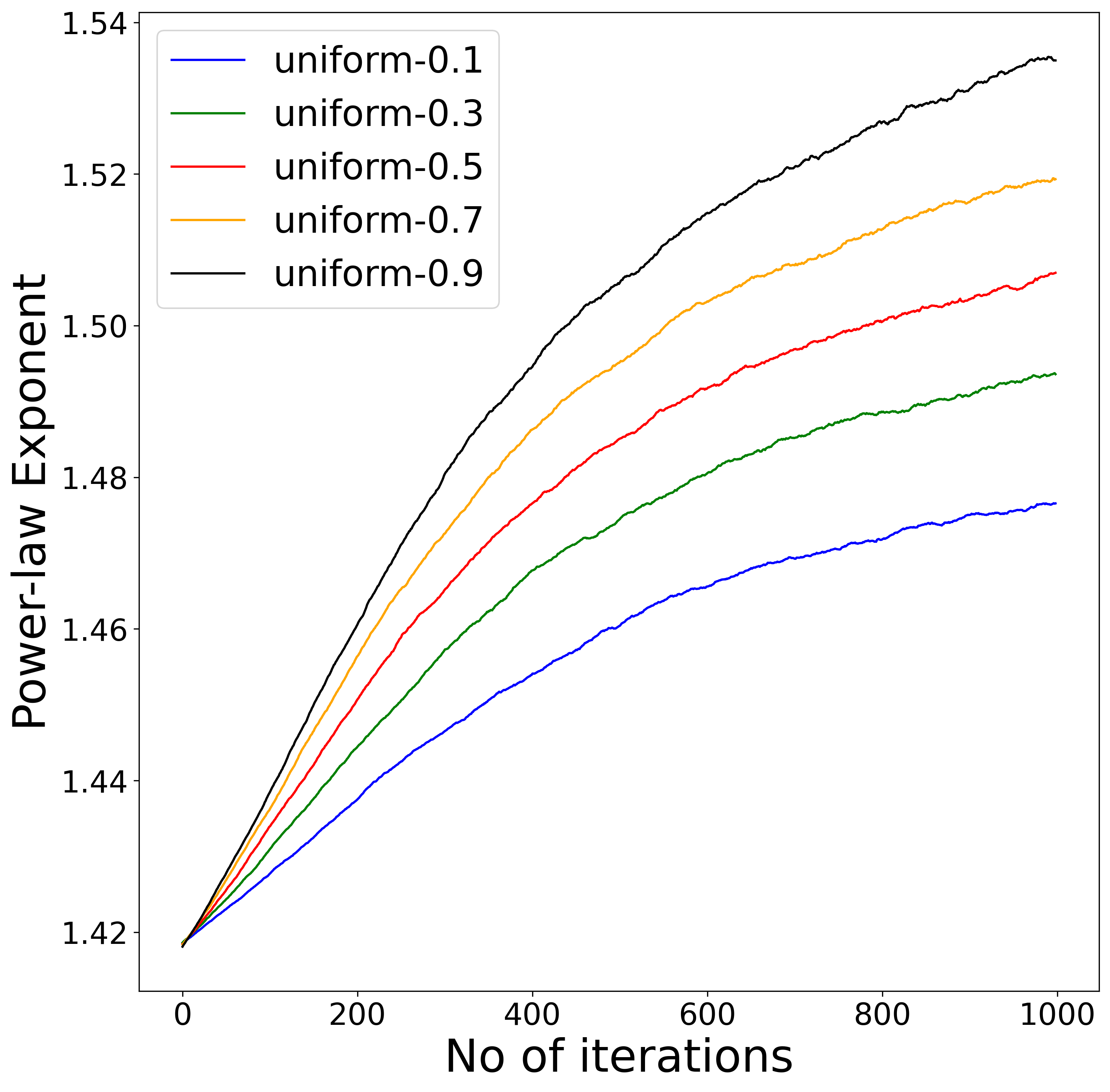}
        \caption{Polynomial}
        \label{uniform_Powerlaw_2}
    \end{subfigure}
    \caption{Power-law correlation under uniform learning rates.}
    \label{uniform_Powerlaw}
\end{figure*}

\FloatBarrier

\begin{figure*}[htbp]
    \centering
    \begin{subfigure}[t]{0.45\textwidth}
        \centering
        \includegraphics[width=\linewidth]{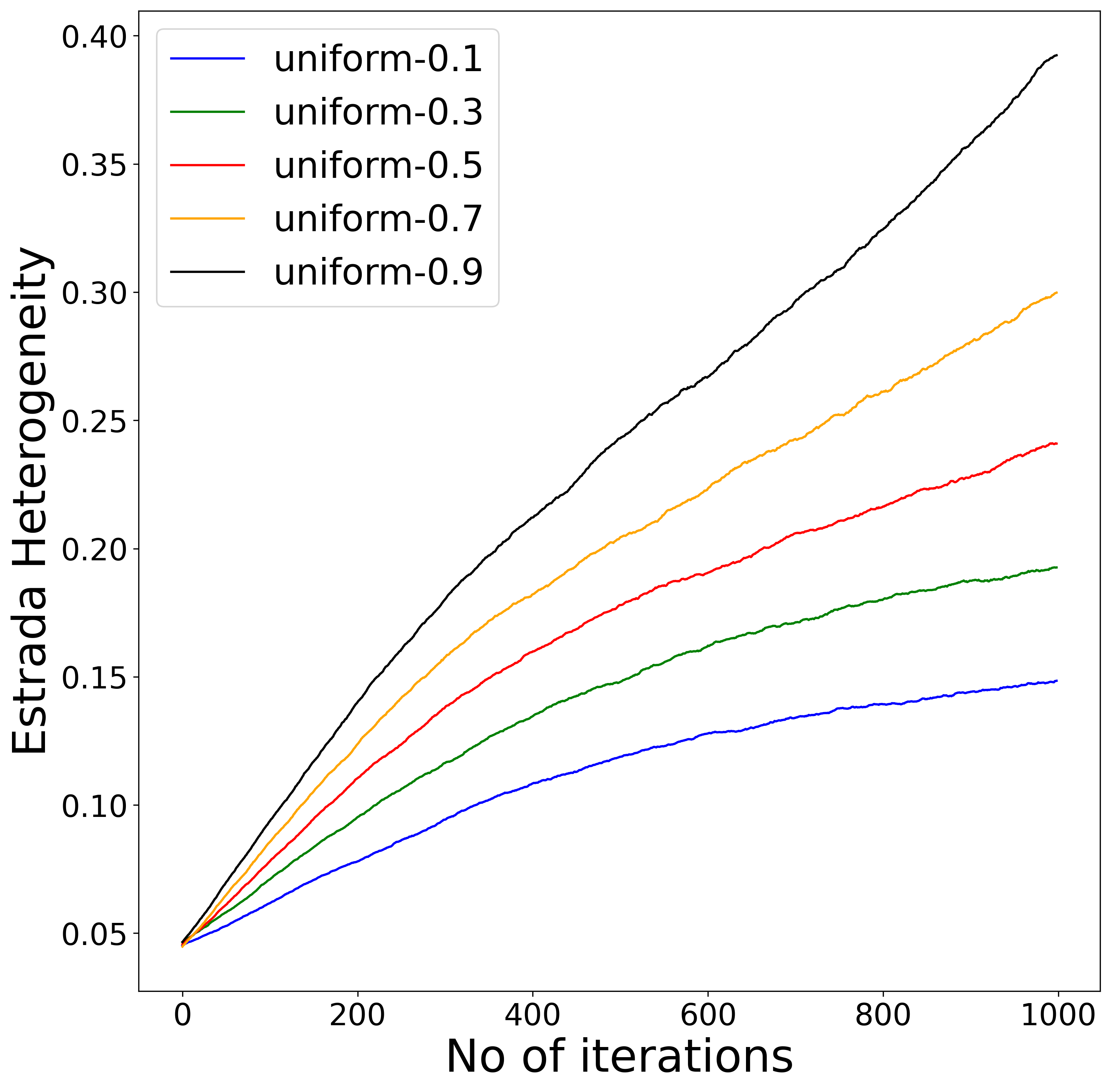}
        \caption{Exponential}
        \label{uniform_Estrada_1}
    \end{subfigure}
    \hfill
    \begin{subfigure}[t]{0.45\textwidth}
        \centering
        \includegraphics[width=\linewidth]{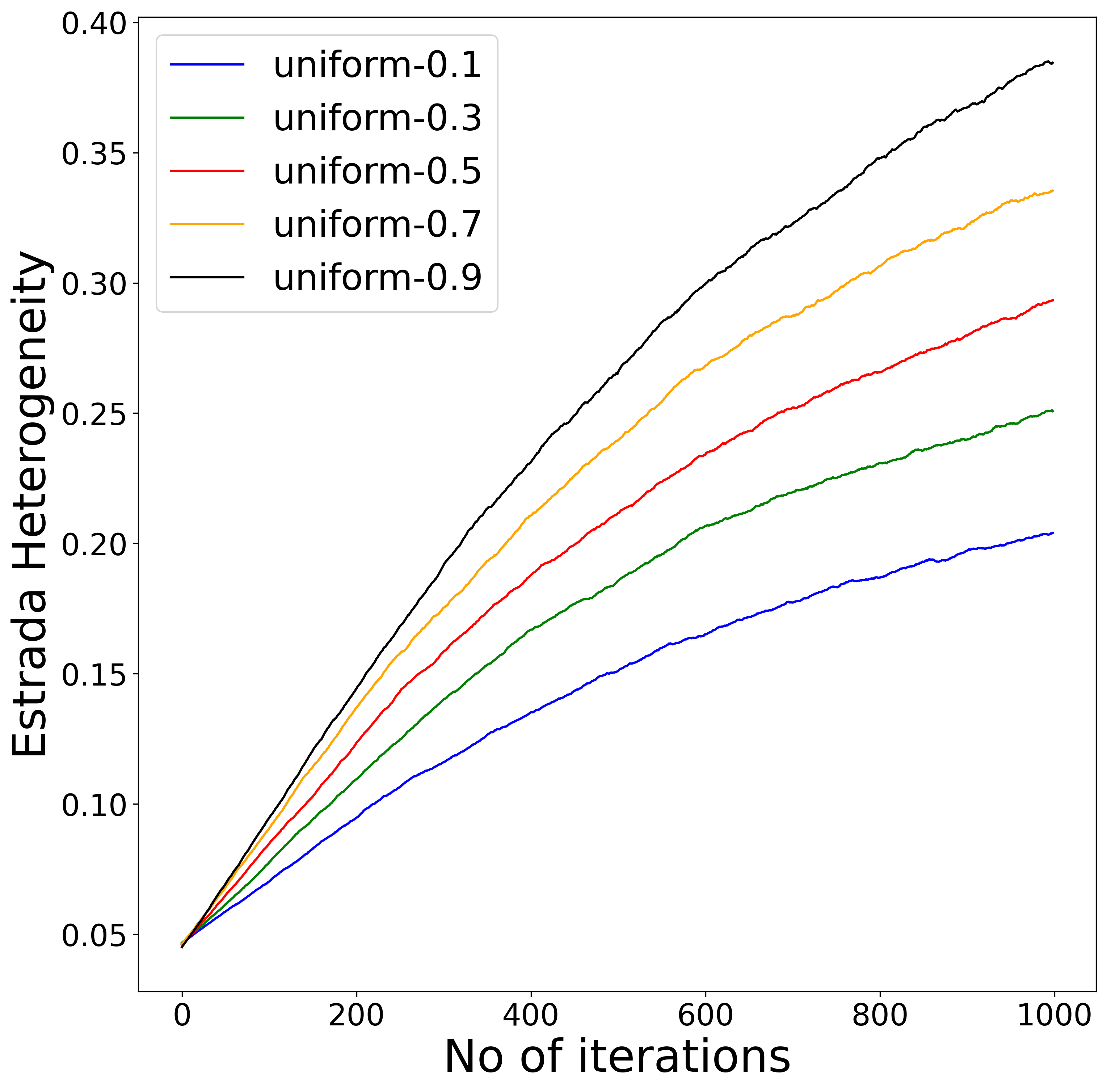}
        \caption{Polynomial}
        \label{uniform_Estrada_2}
    \end{subfigure}
    \caption{Estrada heterogeneity under uniform learning rates.}
    \label{uniform_Estrada}
\end{figure*}

\FloatBarrier

\begin{figure*}[htbp]
    \centering
    \begin{subfigure}[t]{0.45\textwidth}
        \centering
        \includegraphics[width=\linewidth]{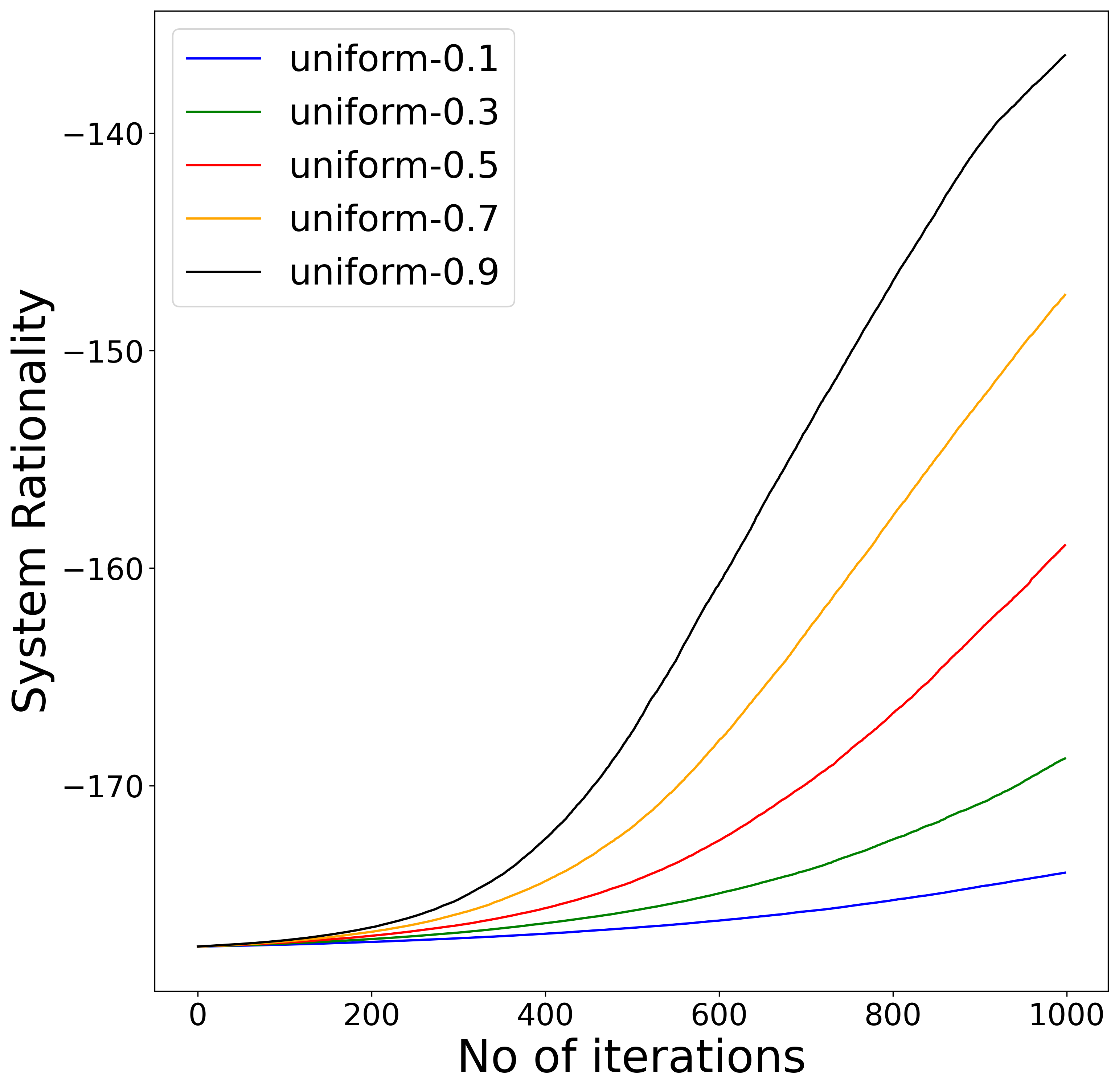}
        \caption{Exponential}
        \label{uniform_System_R_1}
    \end{subfigure}
    \hfill
    \begin{subfigure}[t]{0.45\textwidth}
        \centering
        \includegraphics[width=\linewidth]{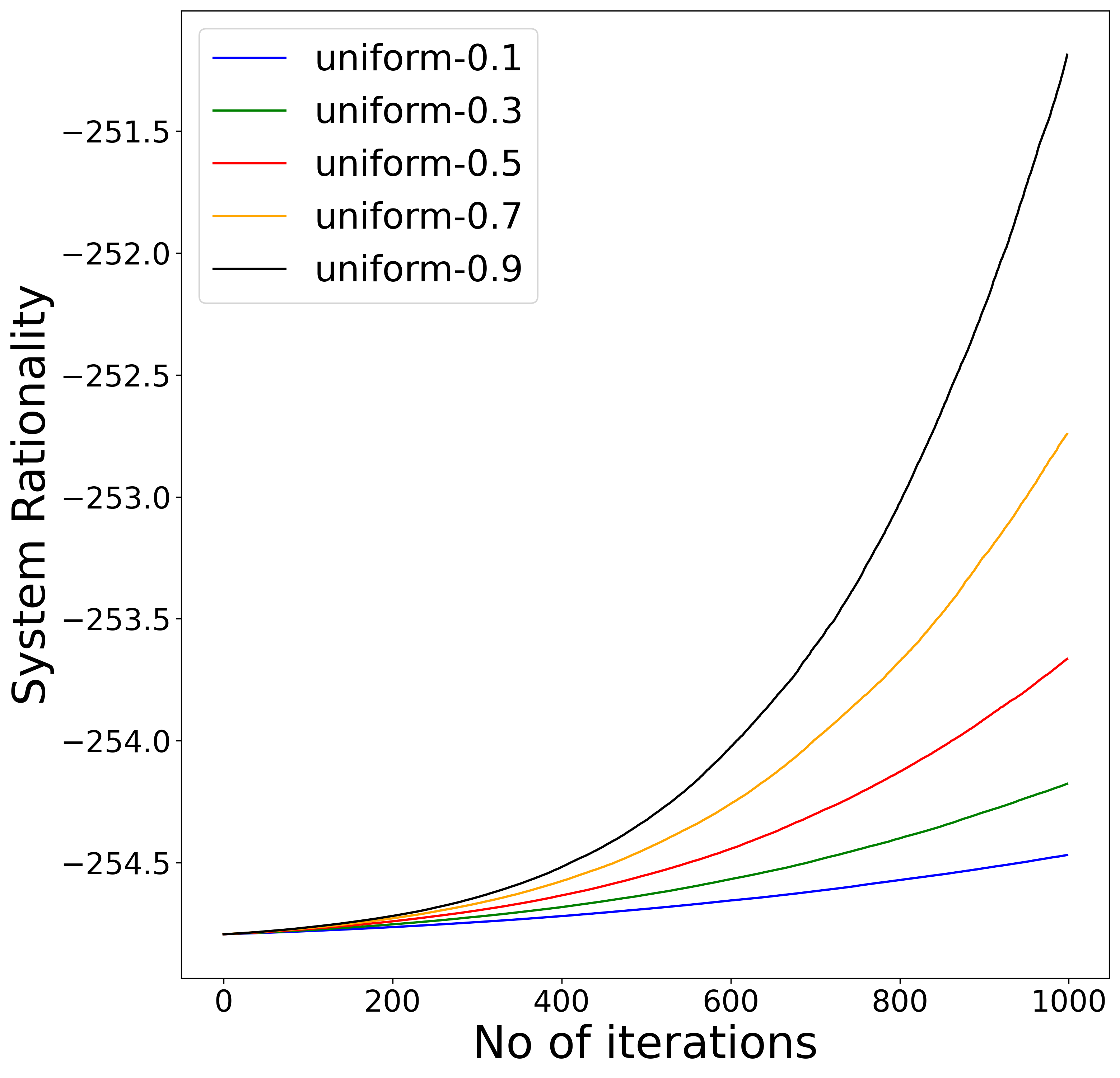}
        \caption{Polynomial}
        \label{uniform_System_R_2}
    \end{subfigure}
    \caption{System rationality under uniform learning rates.}
    \label{uniform_System_R}
\end{figure*}

\FloatBarrier

\begin{figure*}[htbp]
    \centering
    \begin{subfigure}[t]{0.45\textwidth}
        \centering
        \includegraphics[width=\linewidth]{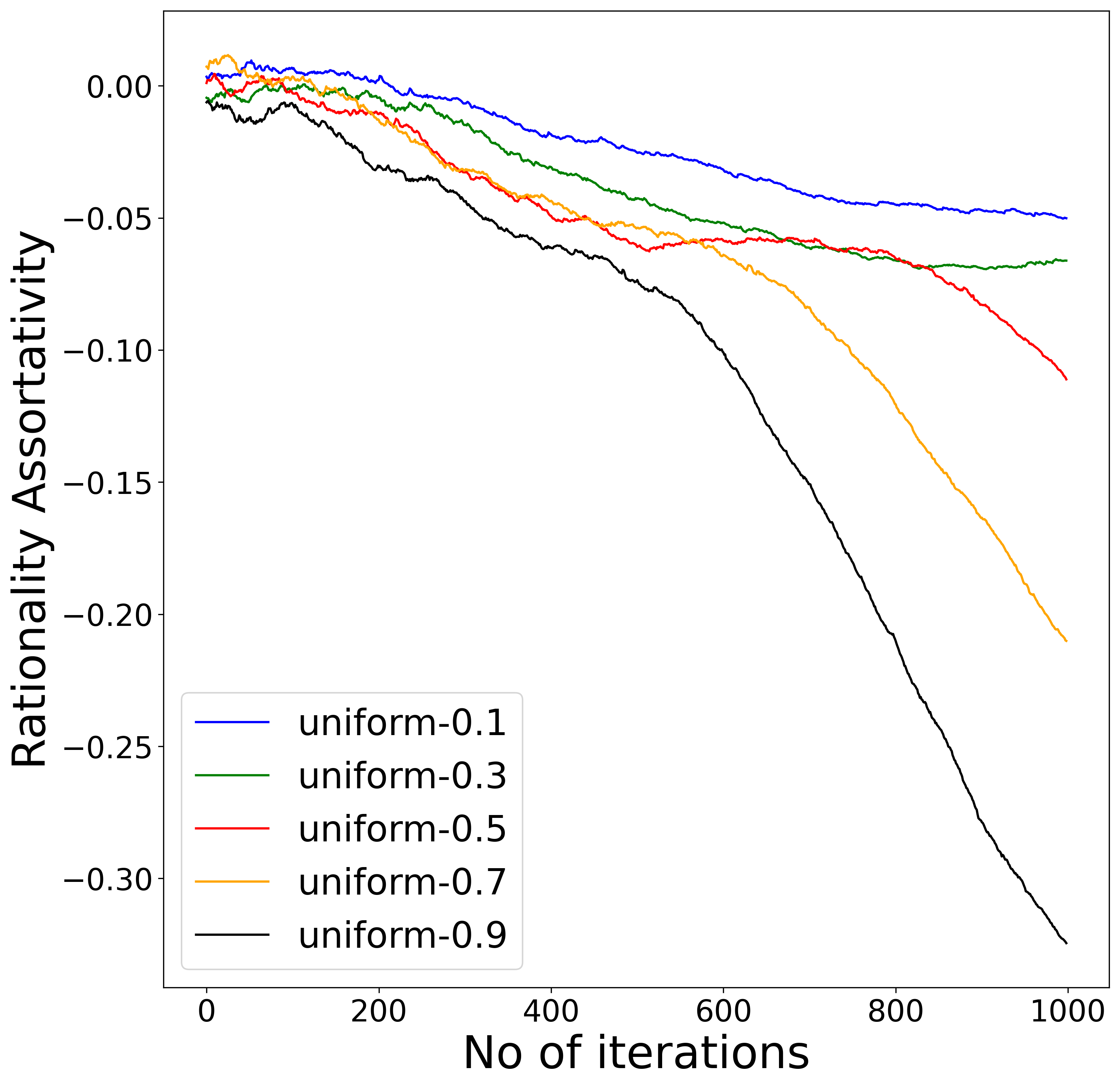}
        \caption{Exponential}
        \label{uniform_R_Assort_1}
    \end{subfigure}
    \hfill
    \begin{subfigure}[t]{0.45\textwidth}
        \centering
        \includegraphics[width=\linewidth]{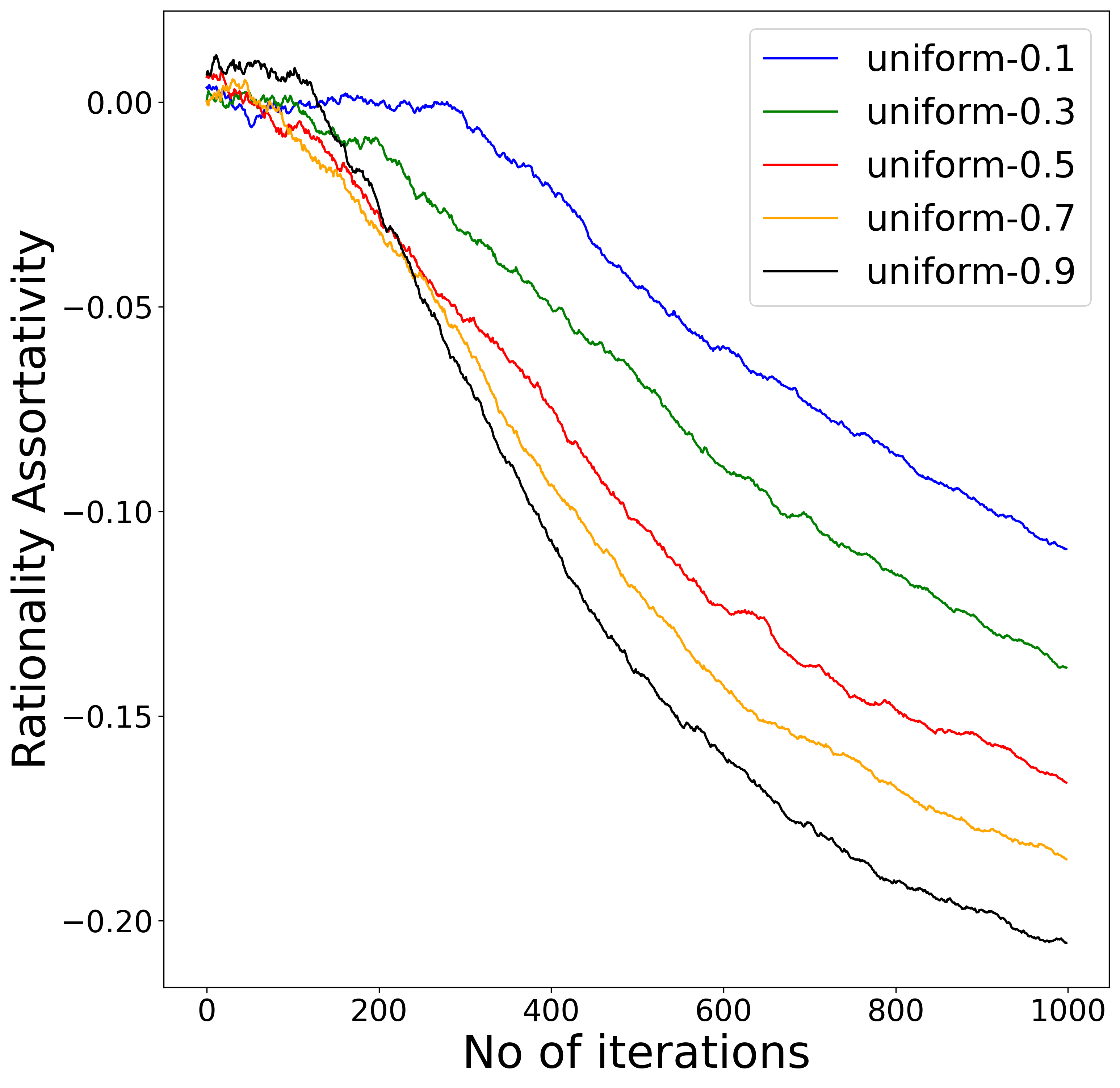}
        \caption{Polynomial}
        \label{uniform_R_Assort_2}
    \end{subfigure}
    \caption{Rationality assortativity under uniform learning rates.}
    \label{uniform_R_Assort}
\end{figure*}

\FloatBarrier

\begin{figure*}[htbp]
    \centering
    \begin{subfigure}[t]{0.45\textwidth}
        \centering
        \includegraphics[width=\linewidth]{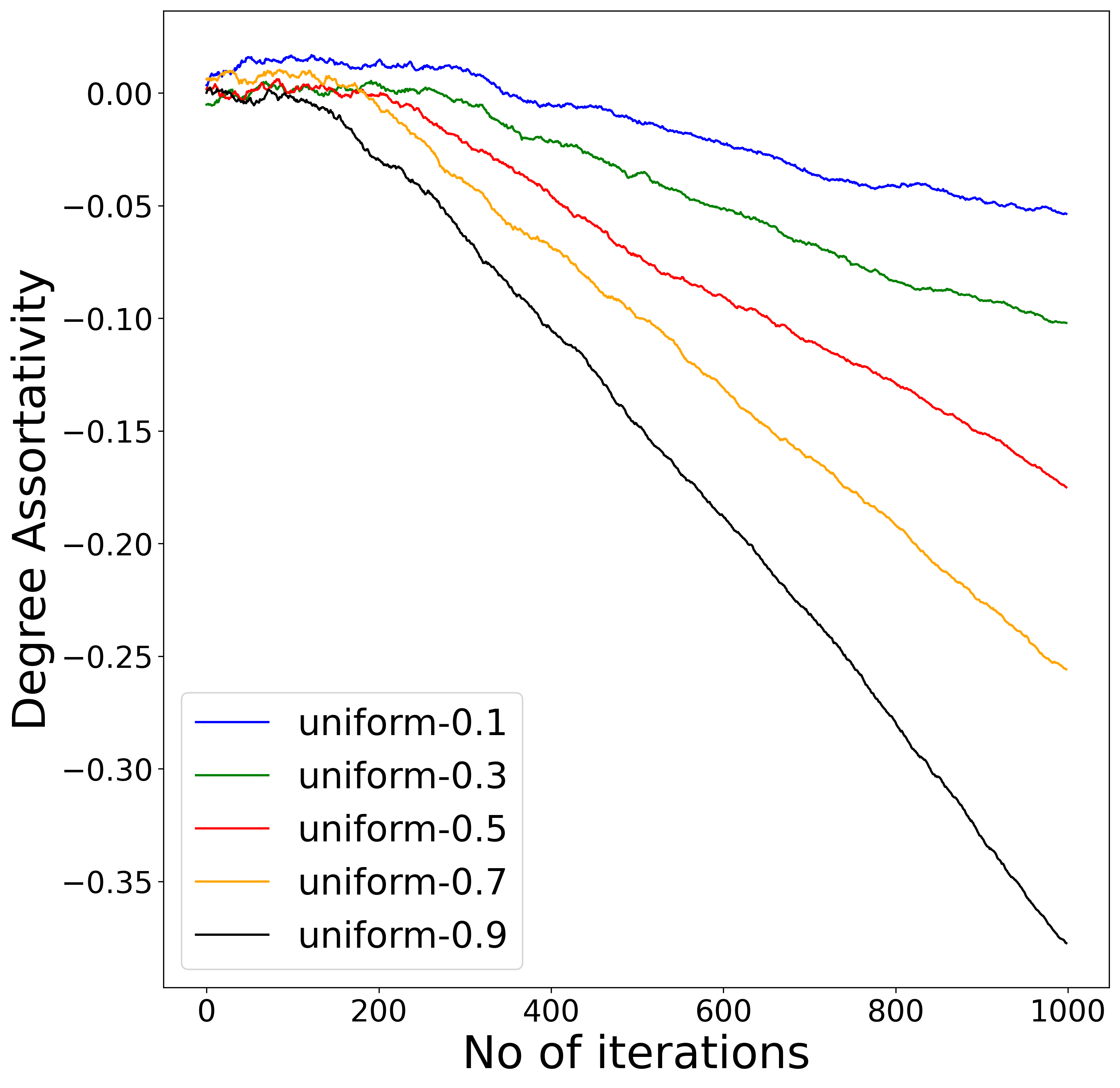}
        \caption{Exponential}
        \label{Uniform_degree_Assort_1}
    \end{subfigure}
    \hfill
    \begin{subfigure}[t]{0.45\textwidth}
        \centering
        \includegraphics[width=\linewidth]{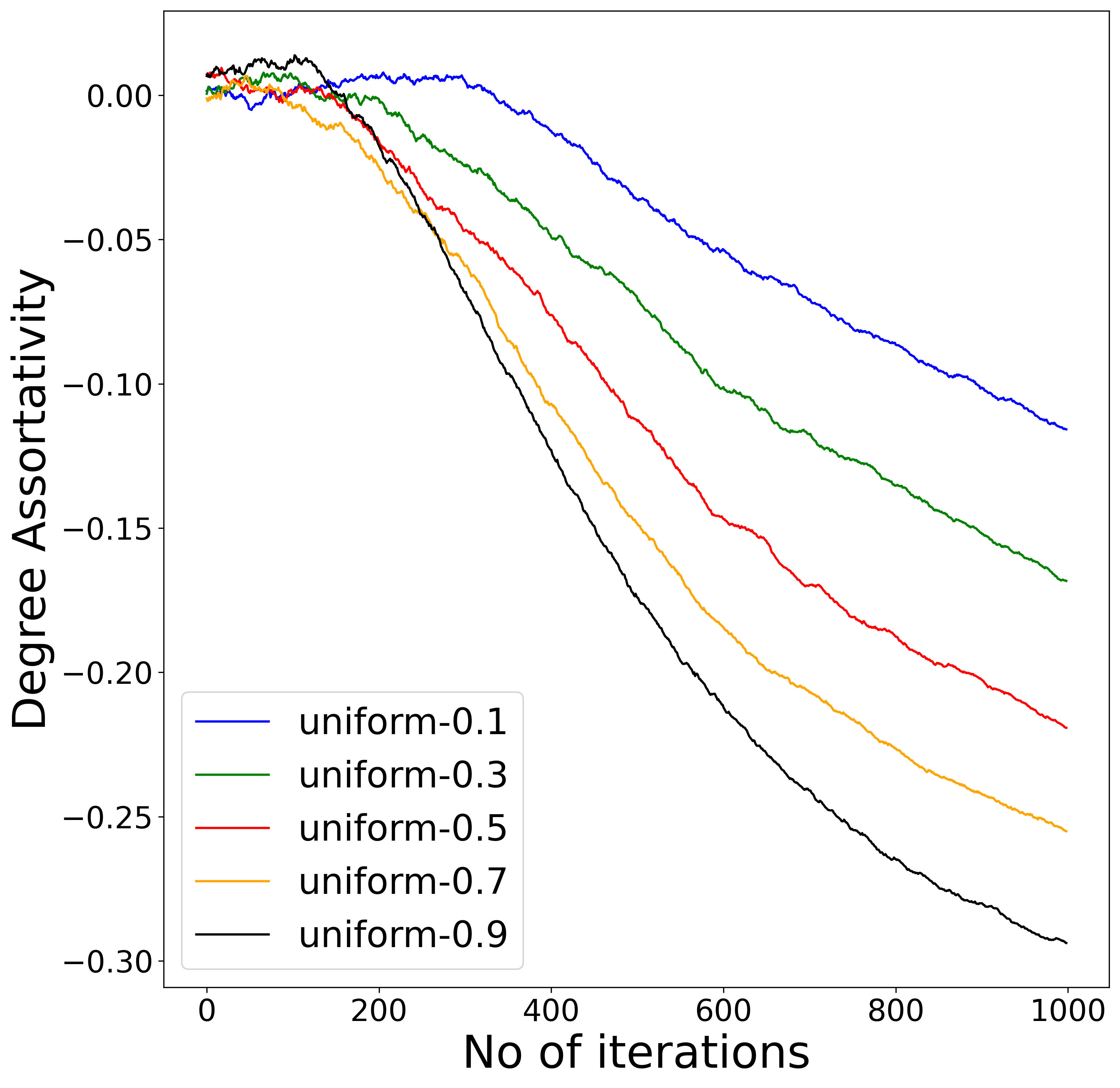}
        \caption{Polynomial}
        \label{Uniform_degree_Assort_2}
    \end{subfigure}
    \caption{Degree assortativity under uniform learning rates.}
    \label{Uniform_degree_Assort}
\end{figure*}

\FloatBarrier

Figures~\ref{uniform_avgR}--\ref{Uniform_degree_Assort} explore the impact of varying uniform learning rates on the topological evolution, when the expected rationality is estimated based on exponential and polynomial formulations, respectively. As shown in Figure~\ref{uniform_avgR}, the average rationality of the nodes increases rapidly under high uniform learning rates, indicating faster convergence toward the expected rationality. Figure~\ref{uniform_Scalefree} confirms the presence of scale-free characteristics under all learning rate settings.

However, the scale-free exponent $\gamma$ varies with the magnitude of the learning rate. For low learning rates, $\gamma$ typically ranges from 1.6 to 1.7. As the learning rate increases, $\gamma$ drops to 1.4--1.5, signifying the formation of core-periphery structures, as shown in Figure~\ref{uniform_Powerlaw}.

The impact on network centralization is also reflected in Figure~\ref{uniform_Estrada}, where the heterogeneity of Estrada increases with the learning rate. This suggests a growing disparity in node communicability, with a few dominant hubs emerging.

The rationality of the system (Figure~\ref{uniform_System_R}) improves with the learning rate as expected from the simulated annealing-based rewiring, but both the rationality assortativity (Figure~\ref{uniform_R_Assort}) and the degree assortativity (Figure~\ref{Uniform_degree_Assort}) decrease, indicating an increase in the disassortative mixing based on degree and node rationality when the population's learning rate increases. 

These findings suggest that lower uniform learning rates result in topologies closer to classical scale-free structures, while higher learning rates drive the network toward a core-periphery topology, as evidenced by a steepening power-law exponent, increased Estrada heterogeneity, and declining assortativity—indicating that rapid adaptation fosters centralization around a few highly rational, well-connected nodes. The results are consistent for both the exponential and polynomial estimates of the expected rationality, suggesting the generalizability of the observations.

\subsection{Learning Rate Variability: Effects of Normal Distribution}
    
\begin{figure*}[htbp]
    \centering
    \begin{subfigure}[t]{0.45\textwidth}
        \centering
        \includegraphics[width=\linewidth]{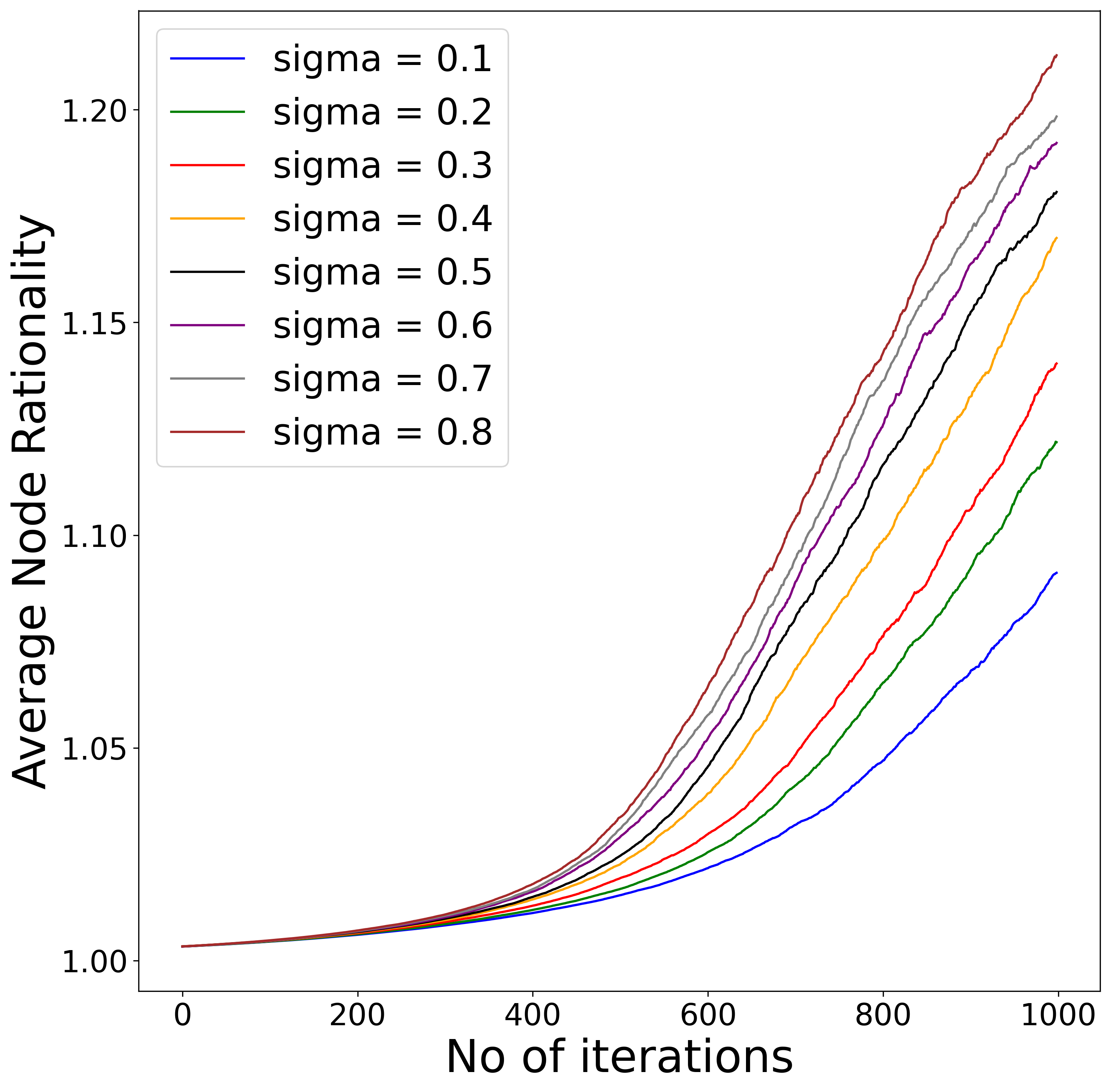}
        \caption{Exponential}
        \label{normal_avg_R_1}
    \end{subfigure}
    \hfill
    \begin{subfigure}[t]{0.45\textwidth}
        \centering
        \includegraphics[width=\linewidth]{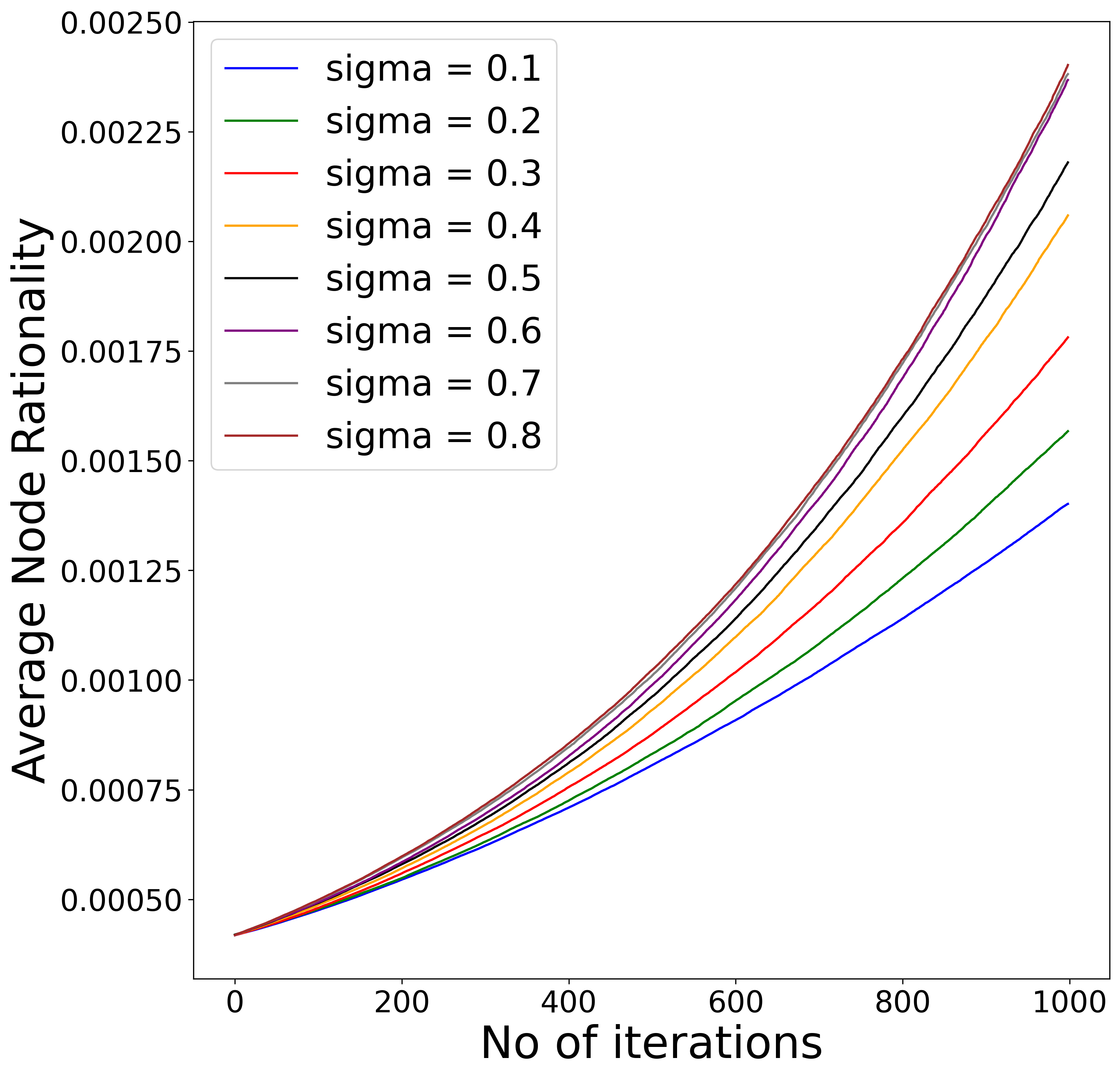}
        \caption{Polynomial}
        \label{normal_avg_R_2}
    \end{subfigure}
    \caption{Average node rationality under different standard deviation values.}
    \label{normal_avg_R}
\end{figure*}

\FloatBarrier
\newpage 
\begin{figure*}[htbp]
    \centering
    \begin{subfigure}[t]{0.45\textwidth}
        \centering
        \includegraphics[width=\linewidth]{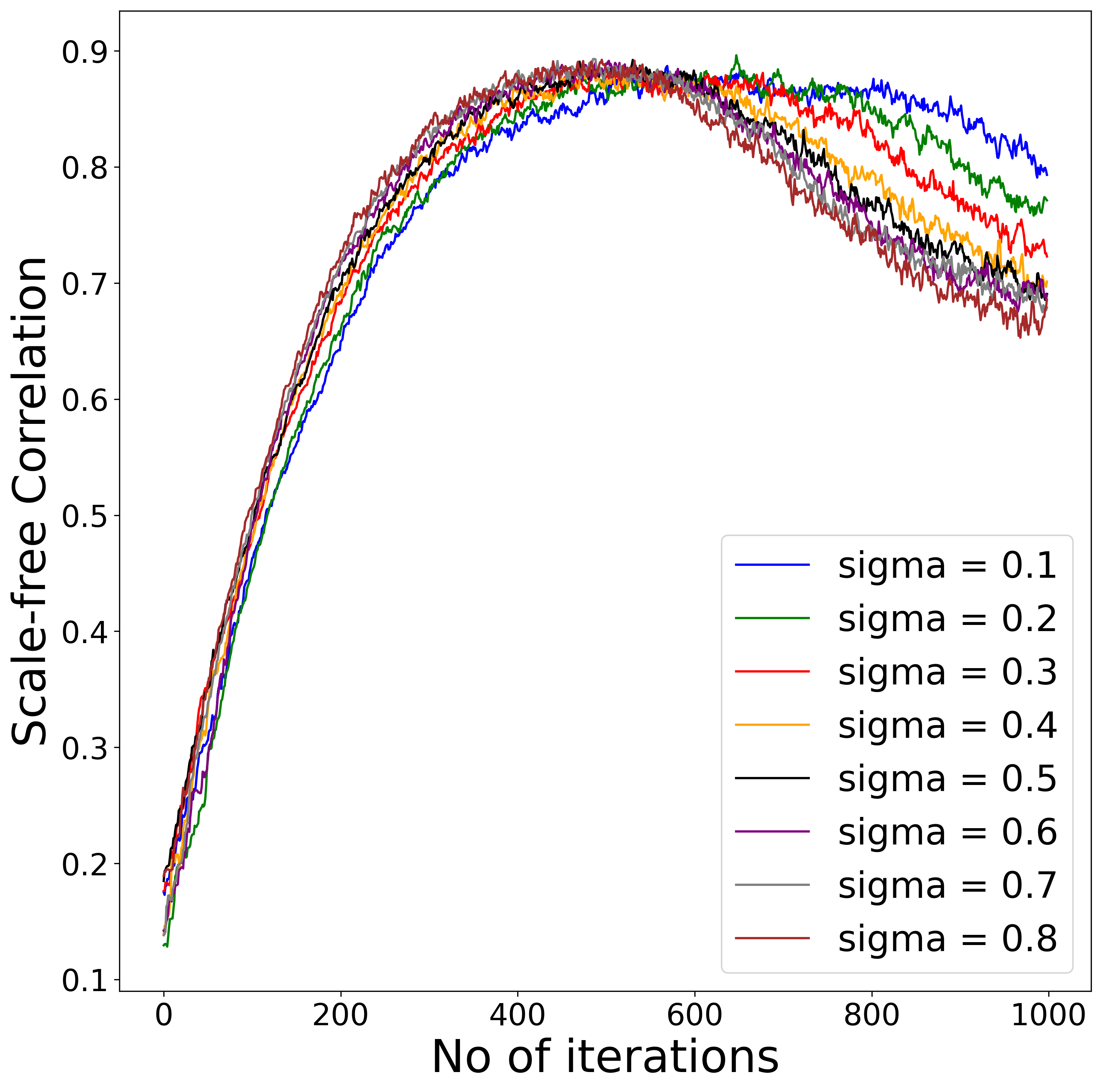}
        \caption{Exponential}
        \label{normal_Scalefree_1}
    \end{subfigure}
    \hfill
    \begin{subfigure}[t]{0.45\textwidth}
        \centering
        \includegraphics[width=\linewidth]{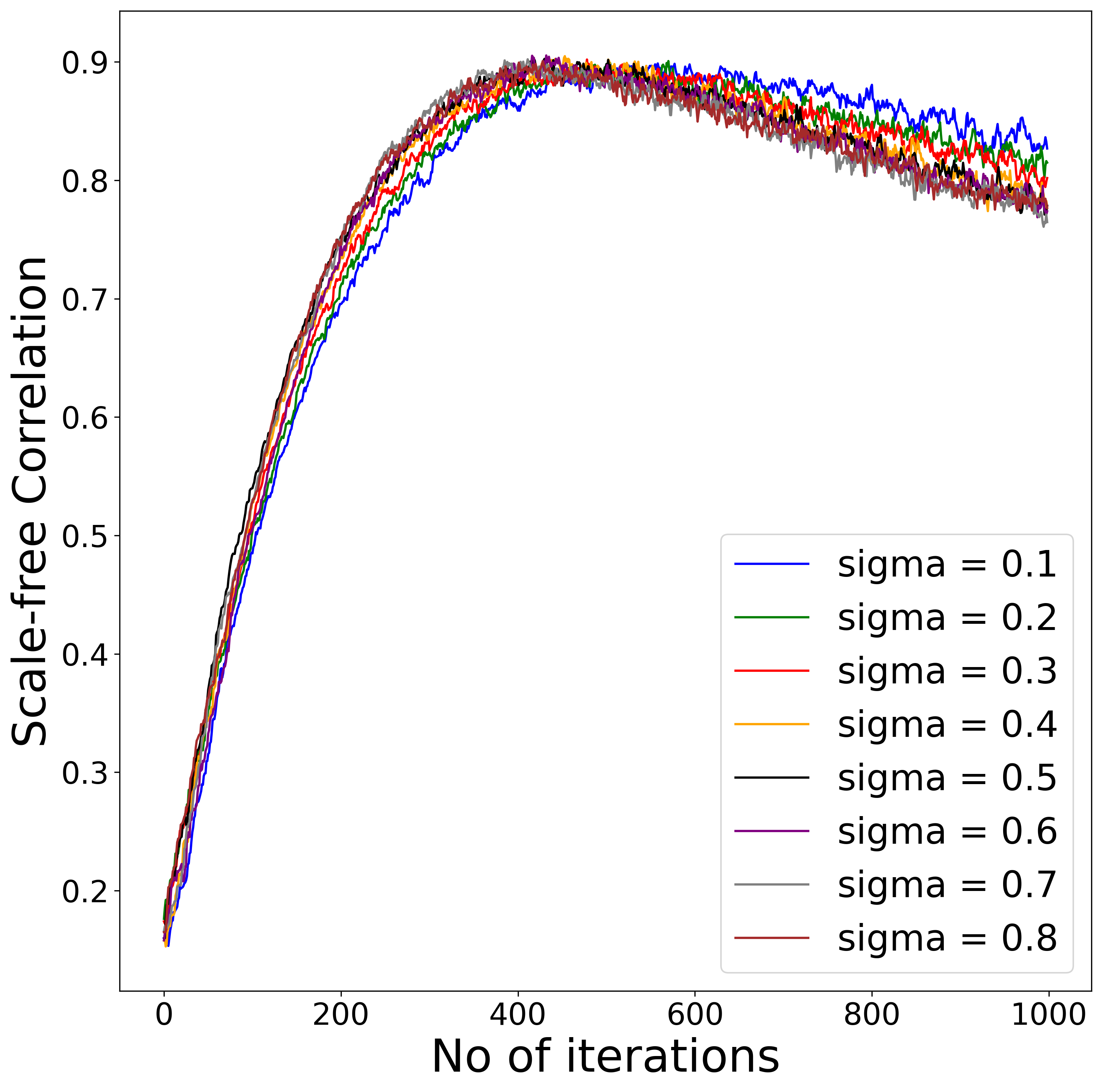}
        \caption{Polynomial}
        \label{normal_Scalefree_2}
    \end{subfigure}
    \caption{Scale-free correlation under different standard deviation values.}
    \label{normal_Scalefree}
\end{figure*}

\FloatBarrier
\begin{figure*}[htbp]
    \centering
    \begin{subfigure}[t]{0.45\textwidth}
        \centering
        \includegraphics[width=\linewidth]{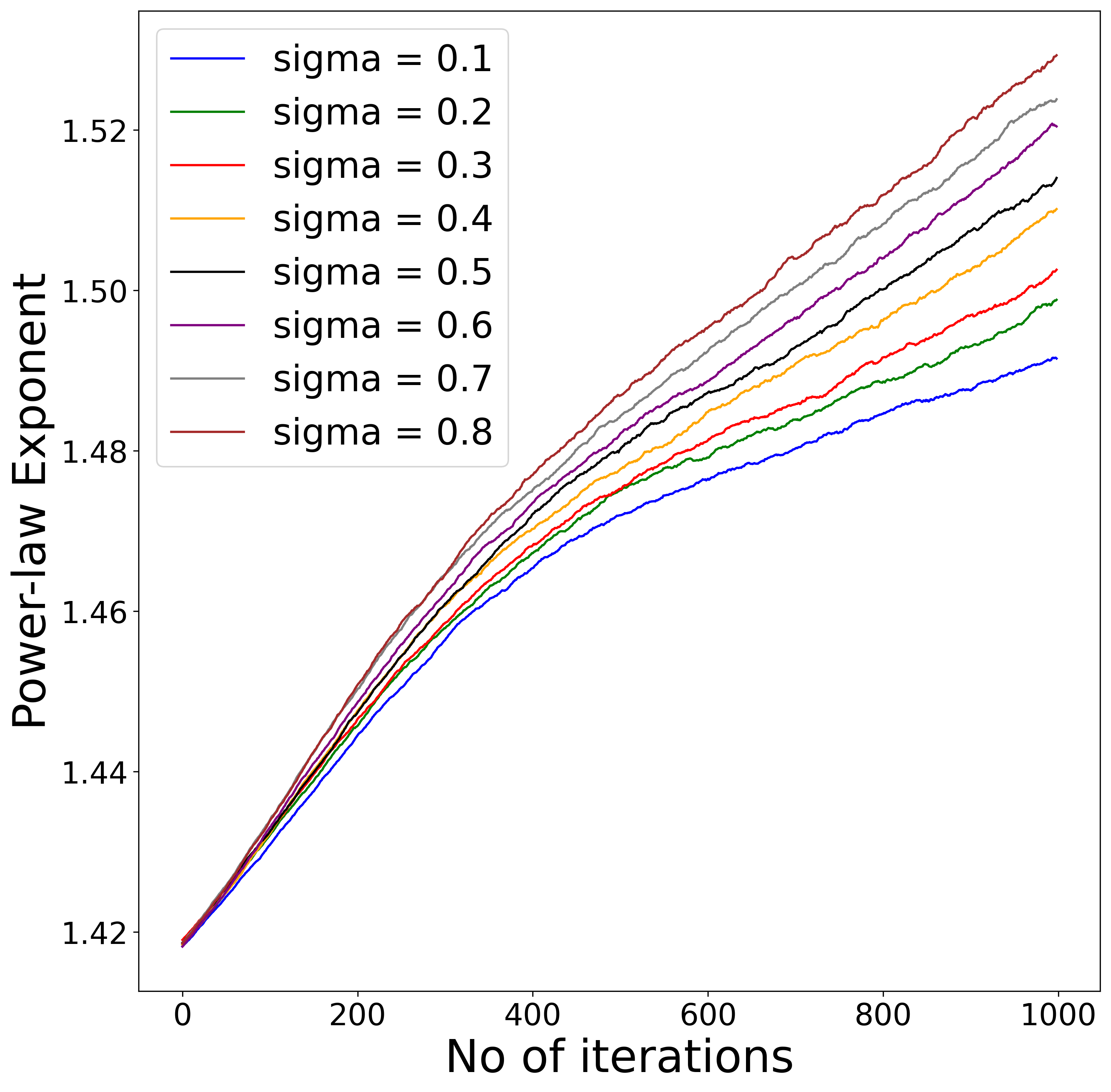}
        \caption{Exponential}
        \label{normal_Powerlaw_1}
    \end{subfigure}
    \hfill
    \begin{subfigure}[t]{0.45\textwidth}
        \centering
        \includegraphics[width=\linewidth]{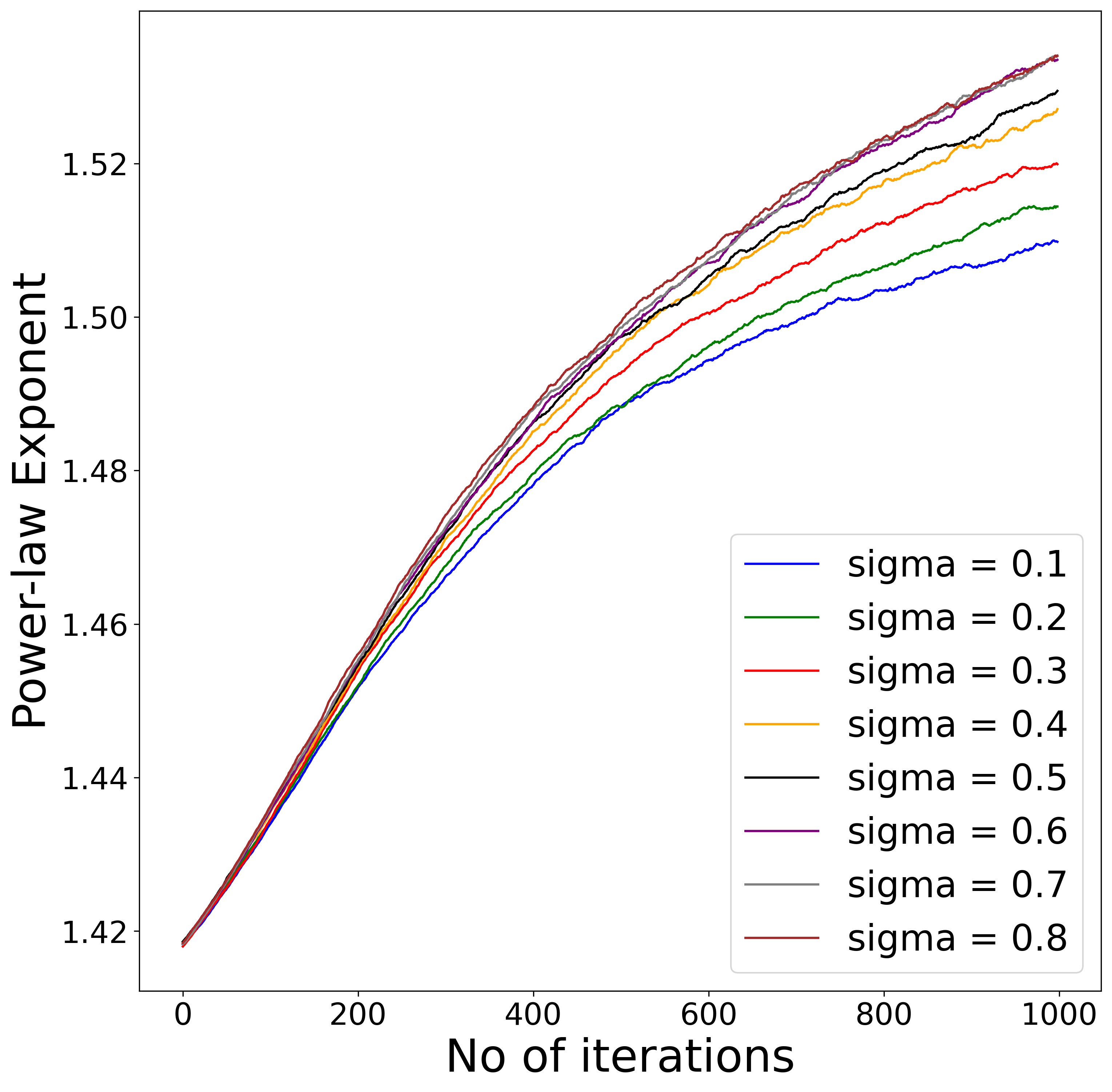}
        \caption{Polynomial}
        \label{normal_Powerlaw_2}
    \end{subfigure}
    \caption{Power-law correlation under different standard deviation values.}
    \label{normal_Powerlaw}
\end{figure*}

\FloatBarrier
\begin{figure*}[htbp]
    \centering
    \begin{subfigure}[t]{0.45\textwidth}
        \centering
        \includegraphics[width=\linewidth]{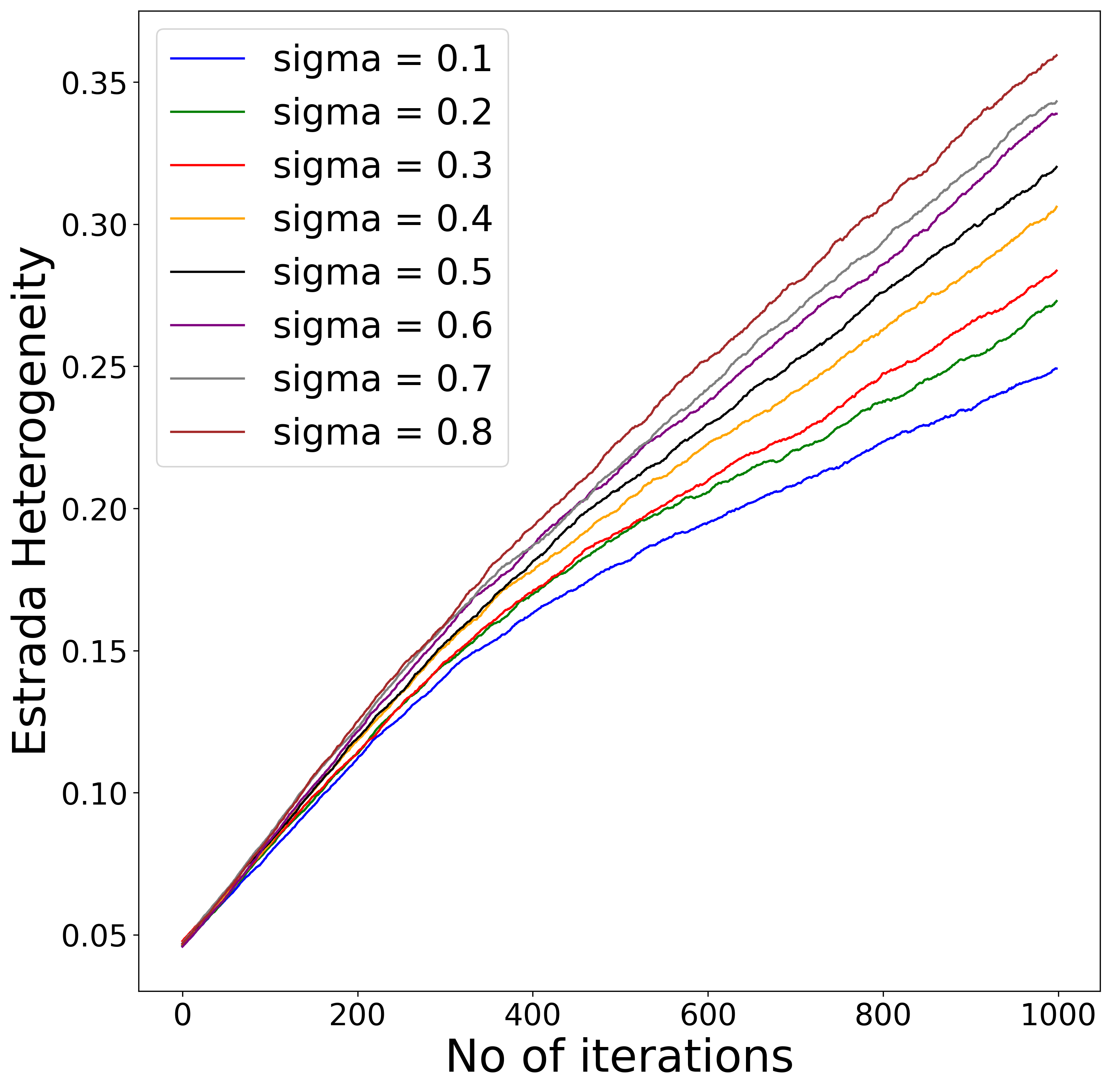}
        \caption{Exponential}
        \label{normal_Estrada_1}
    \end{subfigure}
    \hfill
    \begin{subfigure}[t]{0.45\textwidth}
        \centering
        \includegraphics[width=\linewidth]{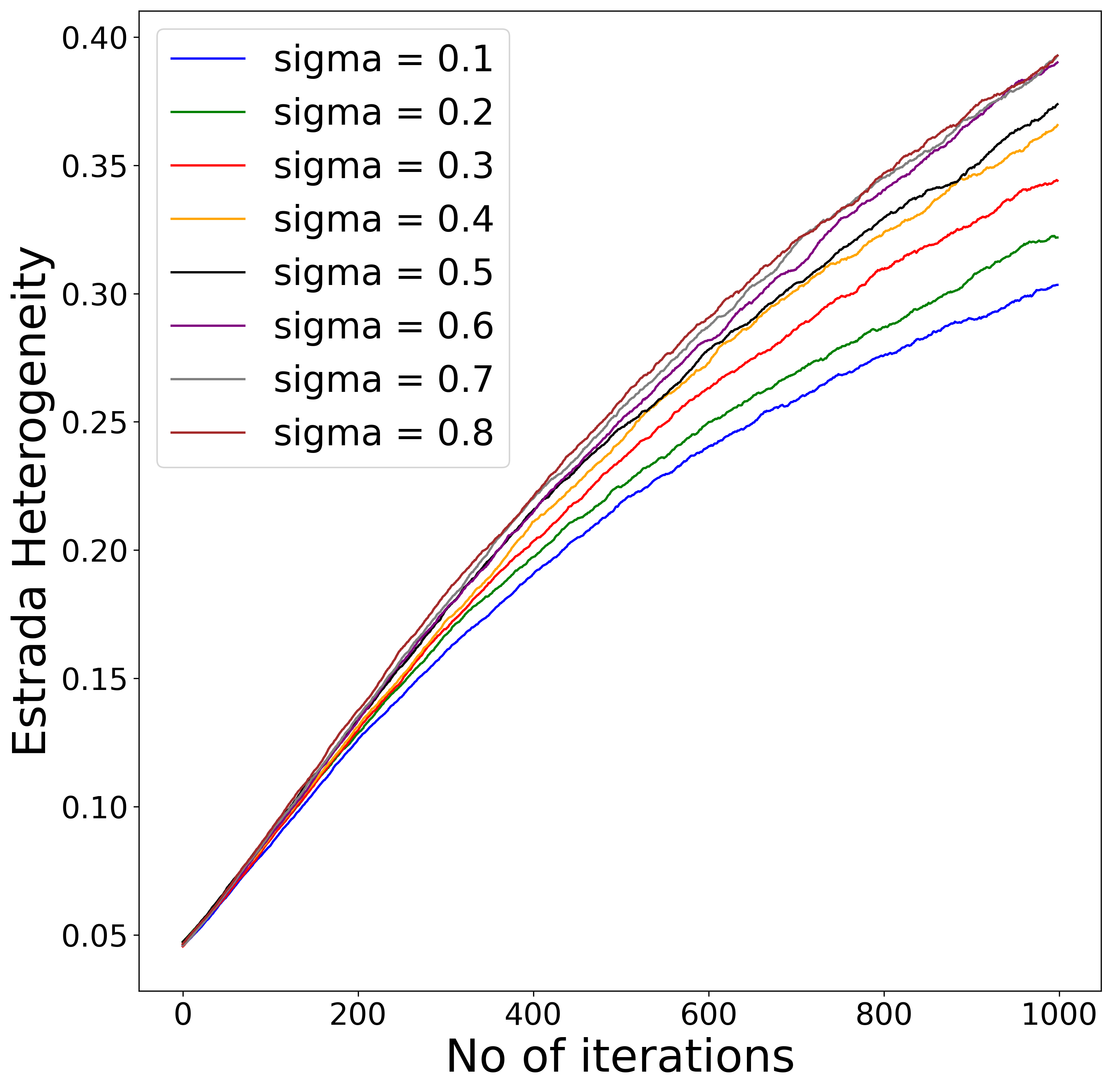}
        \caption{Polynomial}
        \label{normal_Estrada_2}
    \end{subfigure}
    \caption{Estrada heterogeneity under different standard deviation values.}
    \label{normal_Estrada}
\end{figure*}

\FloatBarrier

\begin{figure*}[htbp]
    \centering
    \begin{subfigure}[t]{0.45\textwidth}
        \centering
        \includegraphics[width=\linewidth]{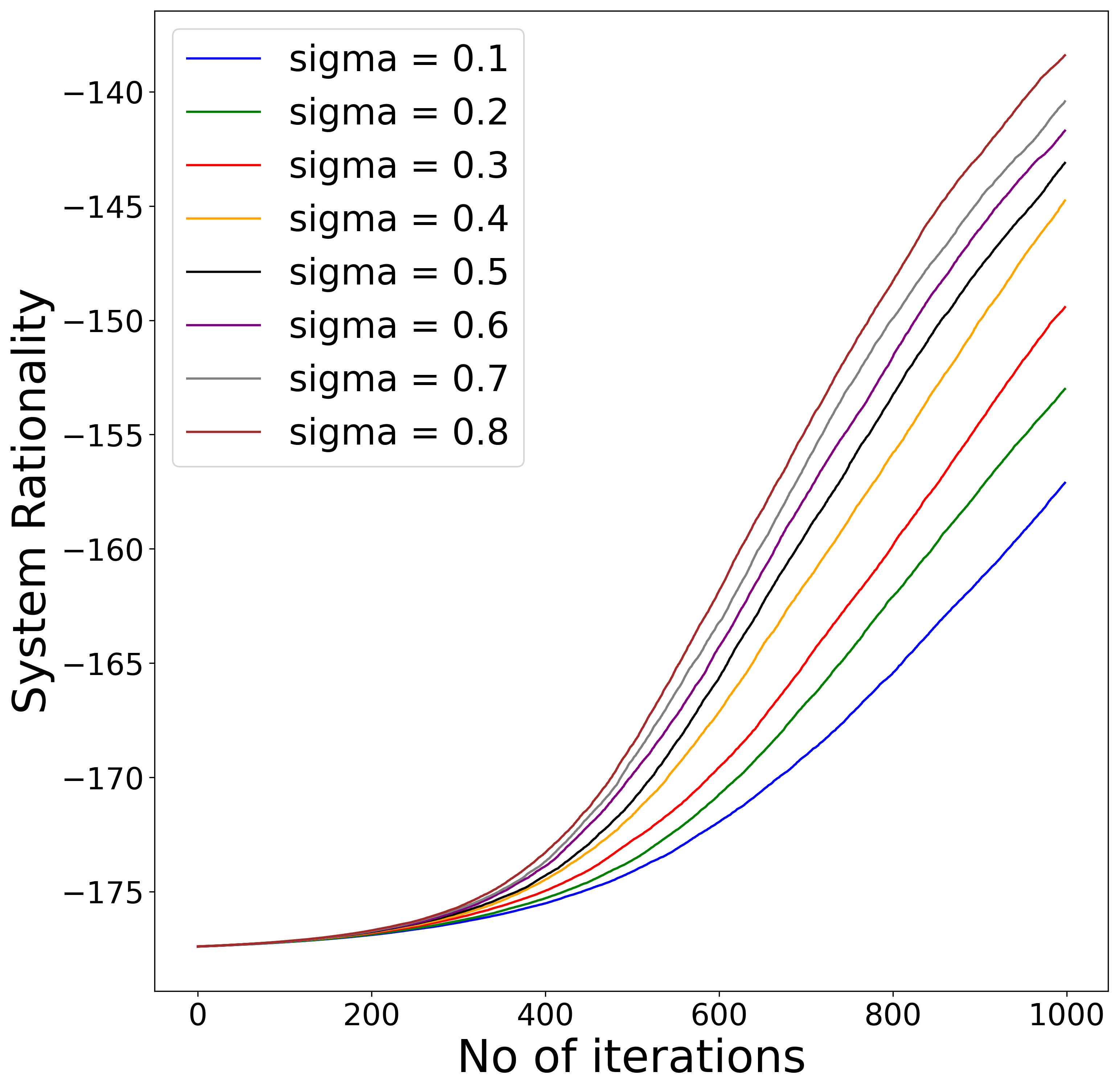}
        \caption{Exponential}
        \label{normal_System_R_1}
    \end{subfigure}
    \hfill
    \begin{subfigure}[t]{0.45\textwidth}
        \centering
        \includegraphics[width=\linewidth]{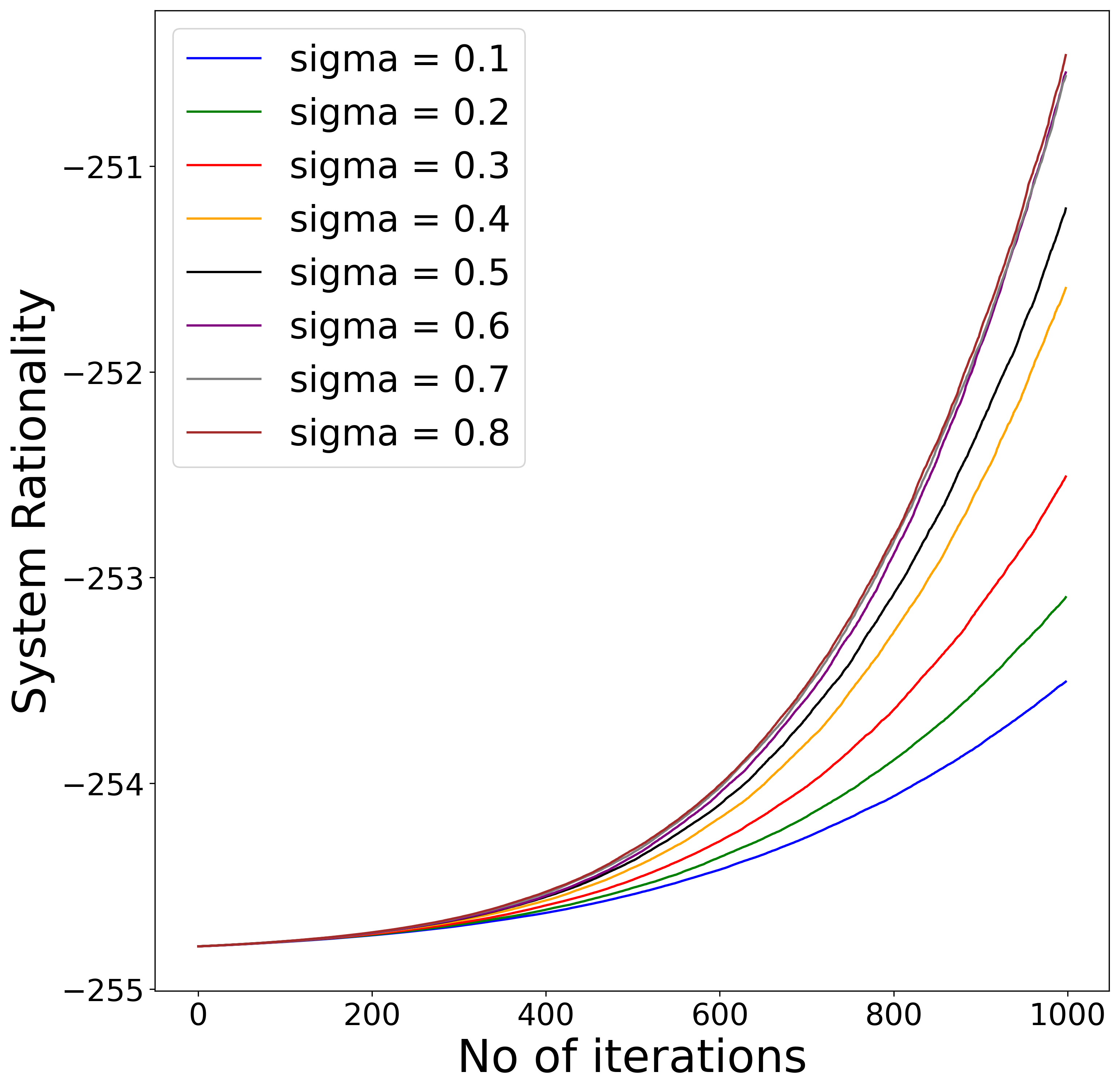}
        \caption{Polynomial}
        \label{normal_System_R_2}
    \end{subfigure}
    \caption{System rationality under different standard deviation values.}
    \label{normal_System_R}
\end{figure*}

\FloatBarrier

\begin{figure*}[htbp]
    \centering
    \begin{subfigure}[t]{0.45\textwidth}
        \centering
        \includegraphics[width=\linewidth]{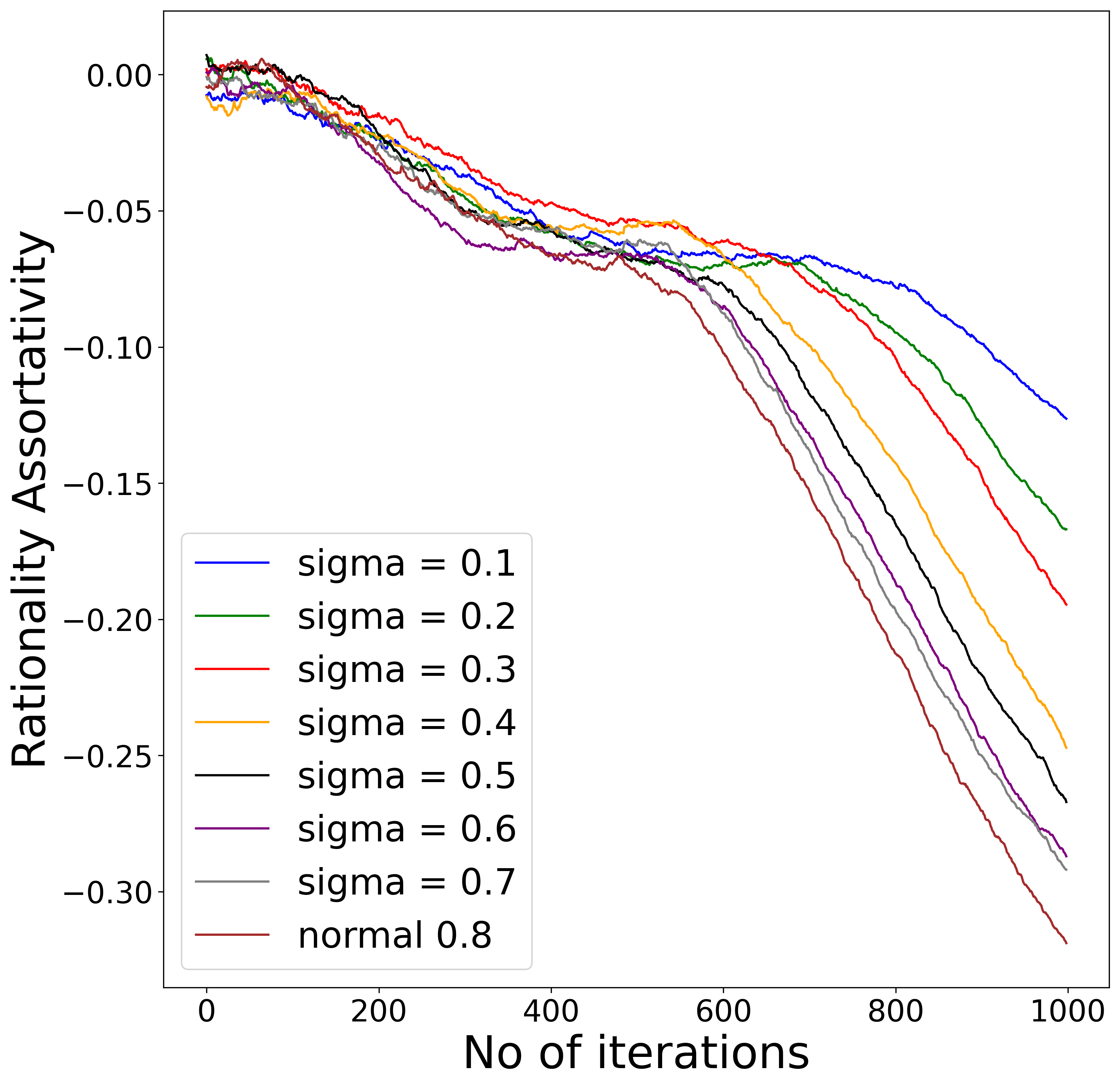}
        \caption{Exponential}
        \label{normal_R_Assort_1}
    \end{subfigure}
    \hfill
    \begin{subfigure}[t]{0.45\textwidth}
        \centering
        \includegraphics[width=\linewidth]{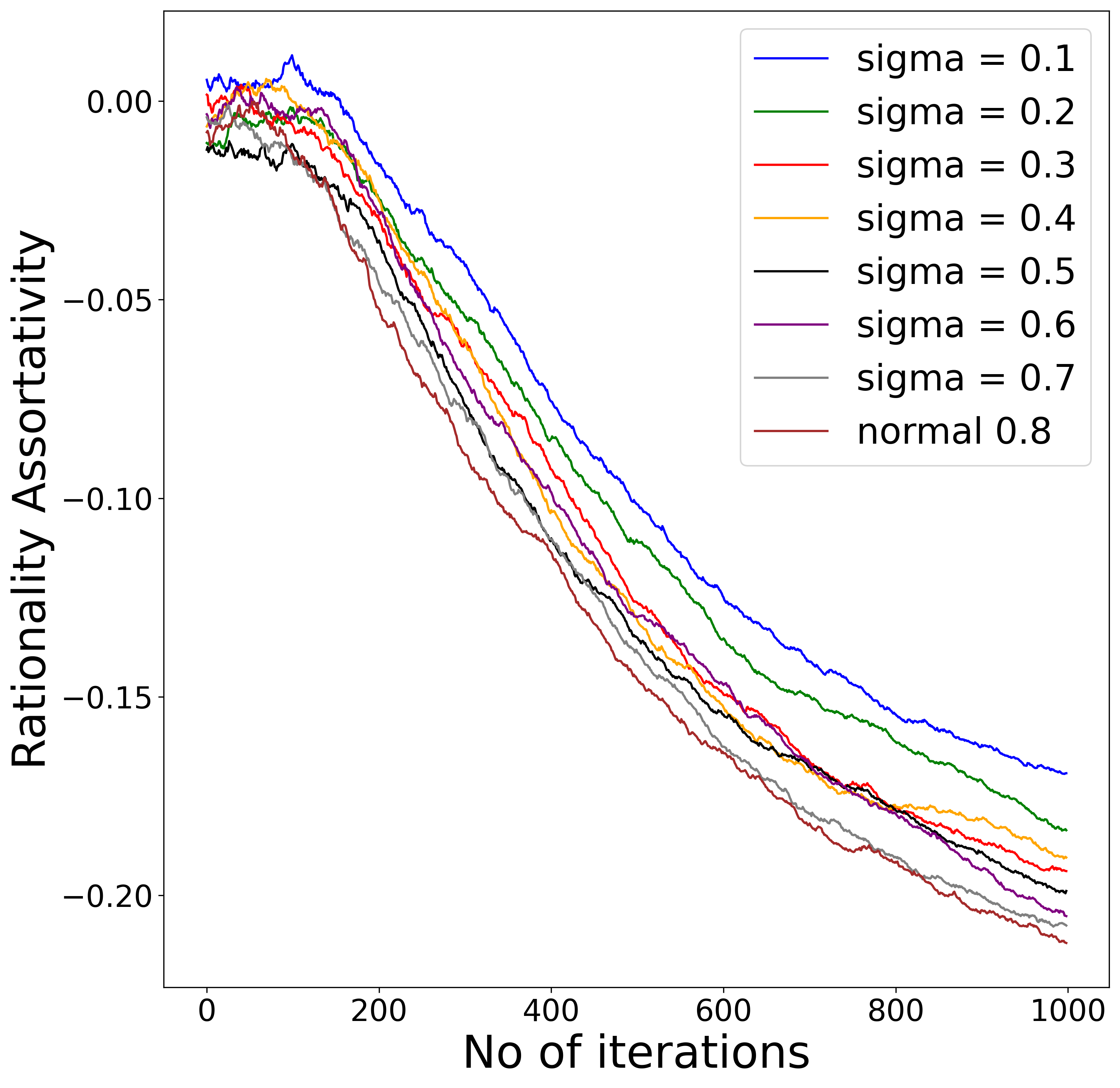}
        \caption{Polynomial}
        \label{normal_R_Assort_2}
    \end{subfigure}
    \caption{Rationality assortativity under different standard deviation values.}
    \label{normal_R_Assort}
\end{figure*}

\FloatBarrier

\begin{figure*}[htbp]
    \centering
    \begin{subfigure}[t]{0.45\textwidth}
        \centering
        \includegraphics[width=\linewidth]{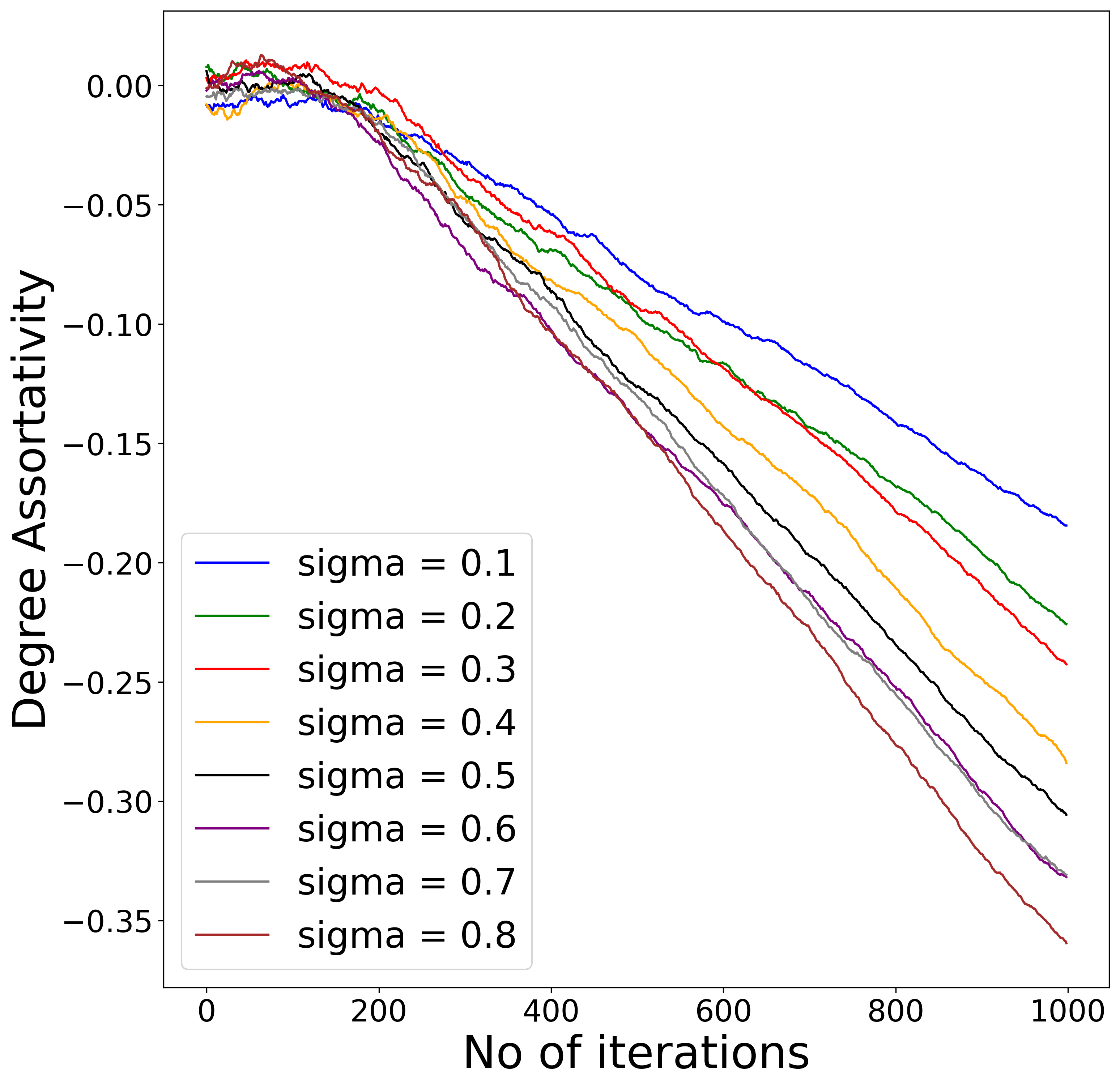}
        \caption{Exponential}
        \label{normal_degree_Assort_1}
    \end{subfigure}
    \hfill
    \begin{subfigure}[t]{0.45\textwidth}
        \centering
        \includegraphics[width=\linewidth]{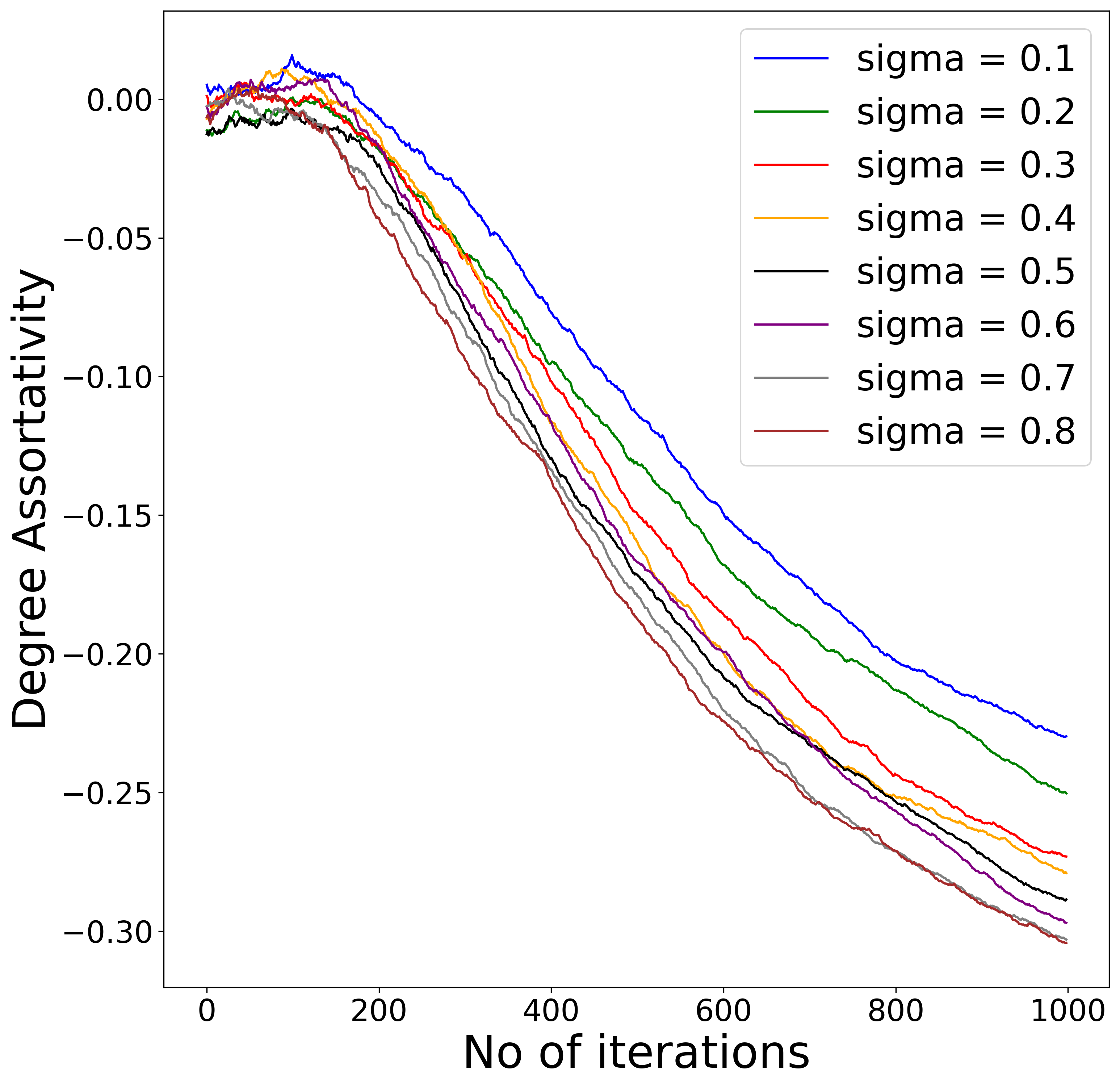}
        \caption{Polynomial}
        \label{normal_degree_Assort_2}
    \end{subfigure}
    \caption{Degree assortativity under different standard deviation values.}
    \label{normal_degree_Assort}
\end{figure*}

\FloatBarrier

Figures~\ref{normal_avg_R}--\ref{normal_degree_Assort} explore the impact of normally distributed learning rates with varying standard deviations on the evolution of topologies. Again, the expected rationality was computed based on both the exponential and polynomial formulations.  

The system maintains strong scale-free correlation (Figure~\ref{normal_Scalefree}) across all settings. However, higher variability leads to emergent heterogeneity, which is evident by increased Estrada heterogeneity in Figure~\ref{normal_Estrada}, when the variability of the rationality increases. Meanwhile, Figure\ref{normal_Powerlaw} shows that the power-law exponent converging to 1.6, even when the standard deviation of the rationality distribution is significantly varied. 

System Rationality improves (Figure~\ref{normal_System_R}) with higher variability of rationality, while Figures~\ref{normal_R_Assort} and \ref{normal_degree_Assort} show decreasing degree assortativity and rationality assortativity, reflecting the increasing role of a few fast-learning agents in shaping global dynamics.

These patterns indicate that increased heterogeneity in learning rates promotes the emergence of core-periphery structures, as evidenced by rising Estrada heterogeneity, a converging power-law exponent near 1.6, and declining assortativity measures—all pointing toward the centralization of influence among a few highly adaptive agents. The results were consistent for both  formulations of expected rationality.

\subsection{Influence of Rewiring Rate on Topology}

\begin{figure}[htbp]
    \centering
    \includegraphics[scale=0.25]{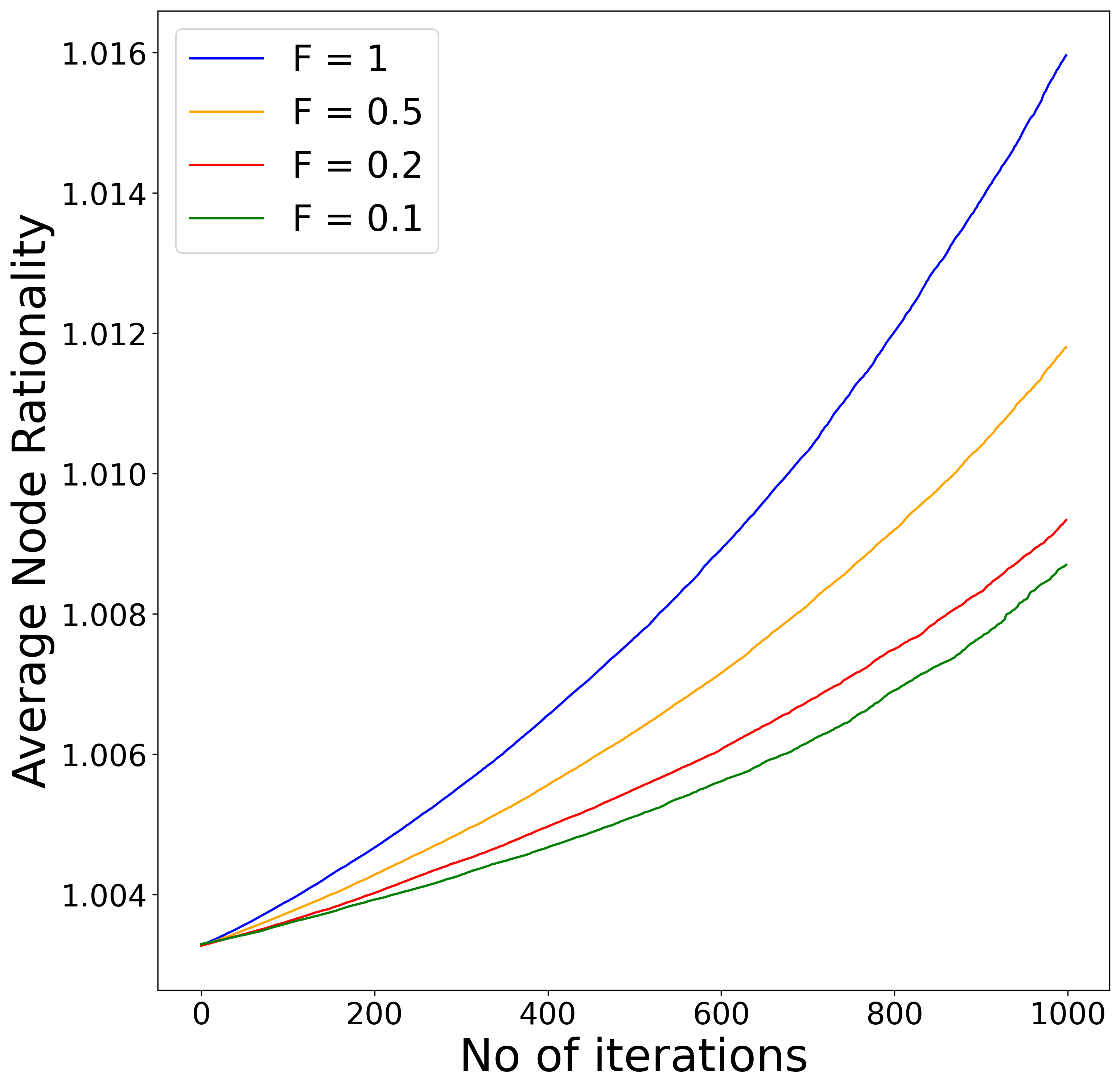}
    \caption{\centering Average Node rationality under different rewiring rates} 
    \label{rewiring_AvgR}
\end{figure} 

\FloatBarrier

\begin{figure}[htbp]
    \centering
    \includegraphics[scale=0.25]{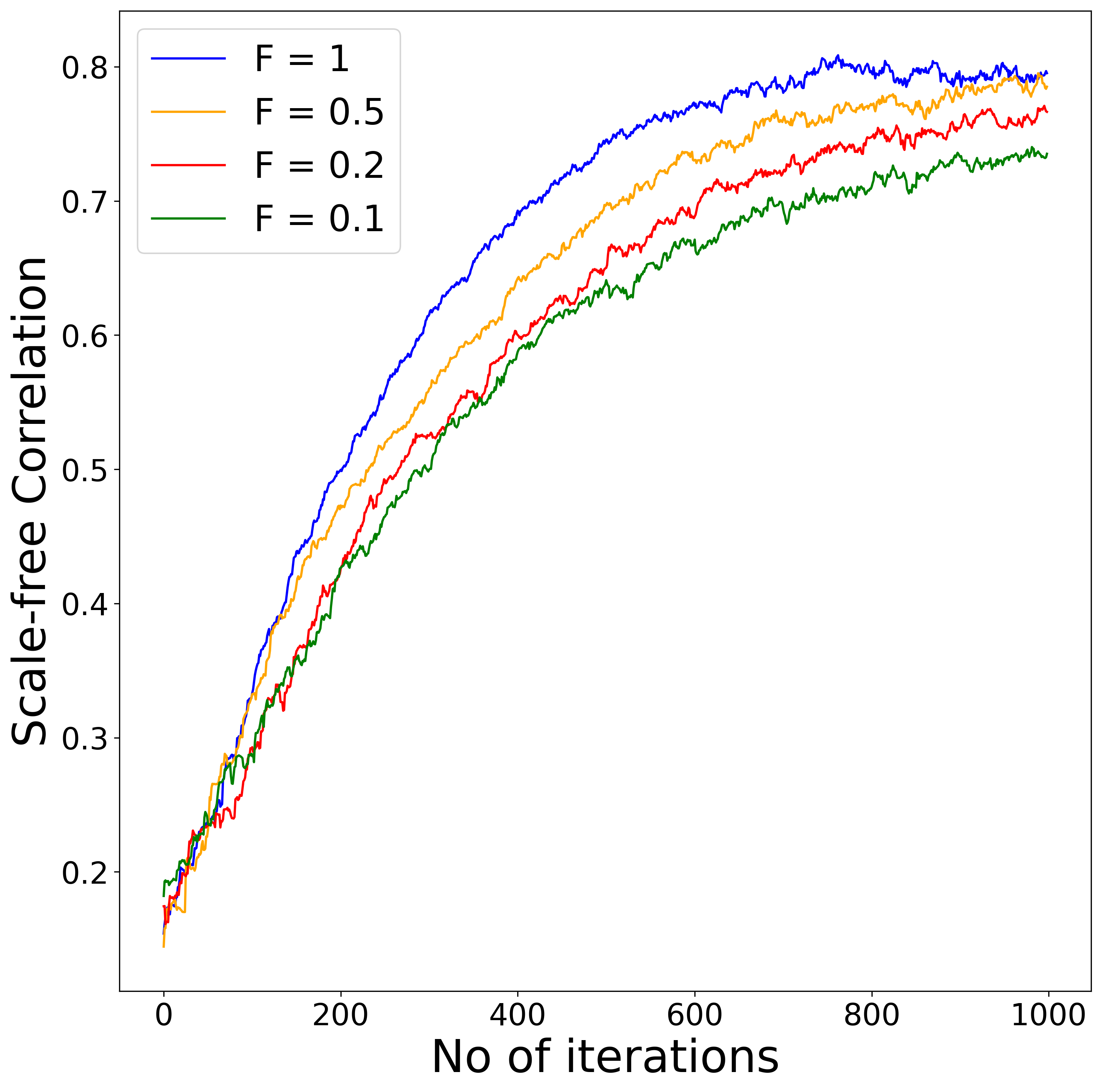}
    \caption{\centering Scale-free Correlation under different rewiring rates} 
    \label{rewiring_Scalefree}
\end{figure} 
\FloatBarrier

\begin{figure}[htbp]
    \centering
    \includegraphics[scale=0.25]{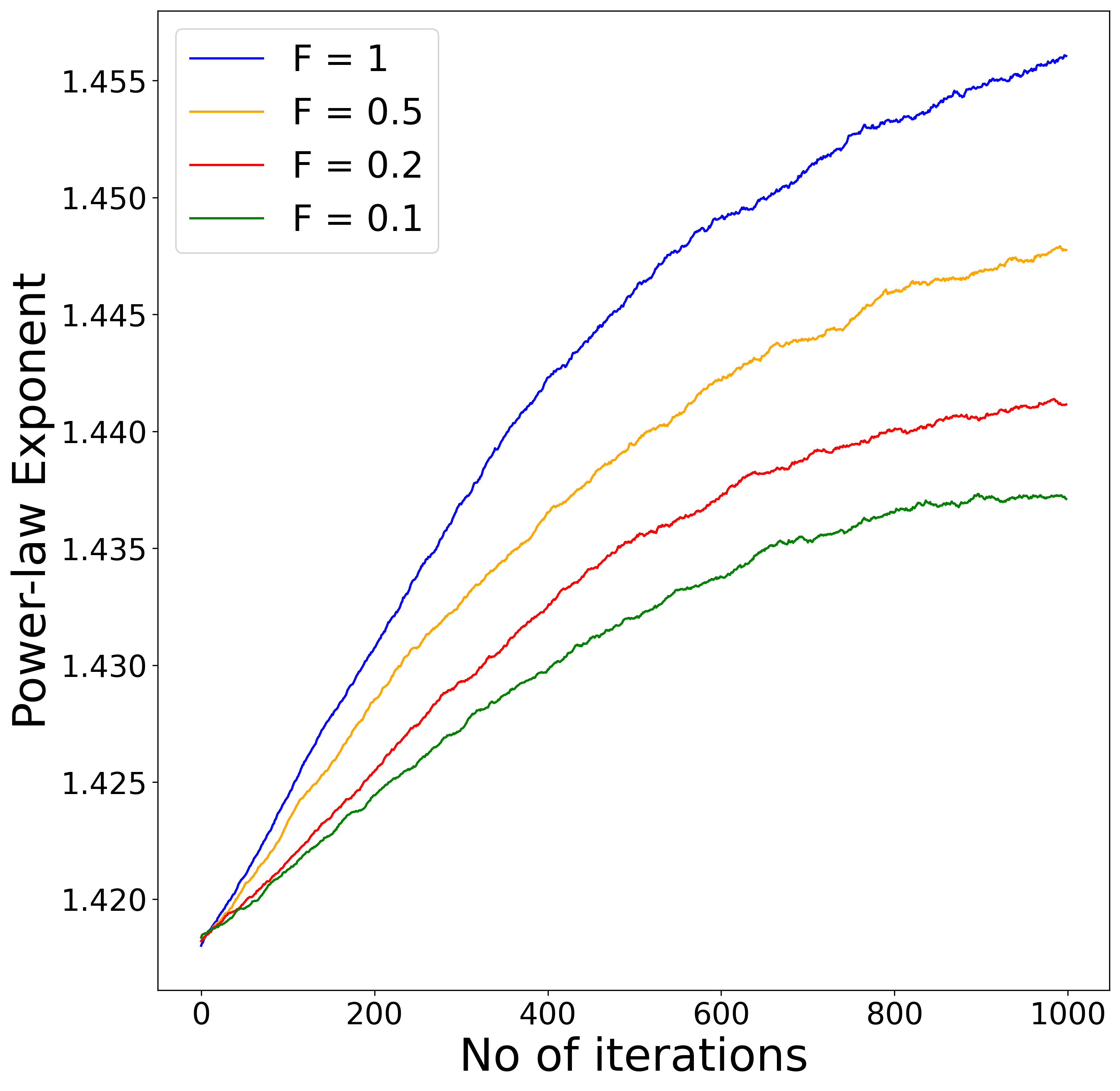}
    \caption{\centering Power-law under different rewiring rates} 
    \label{rewiring_Powerlaw}
\end{figure} 
\FloatBarrier

\begin{figure}[htbp]
    \centering
    \includegraphics[scale=0.25]{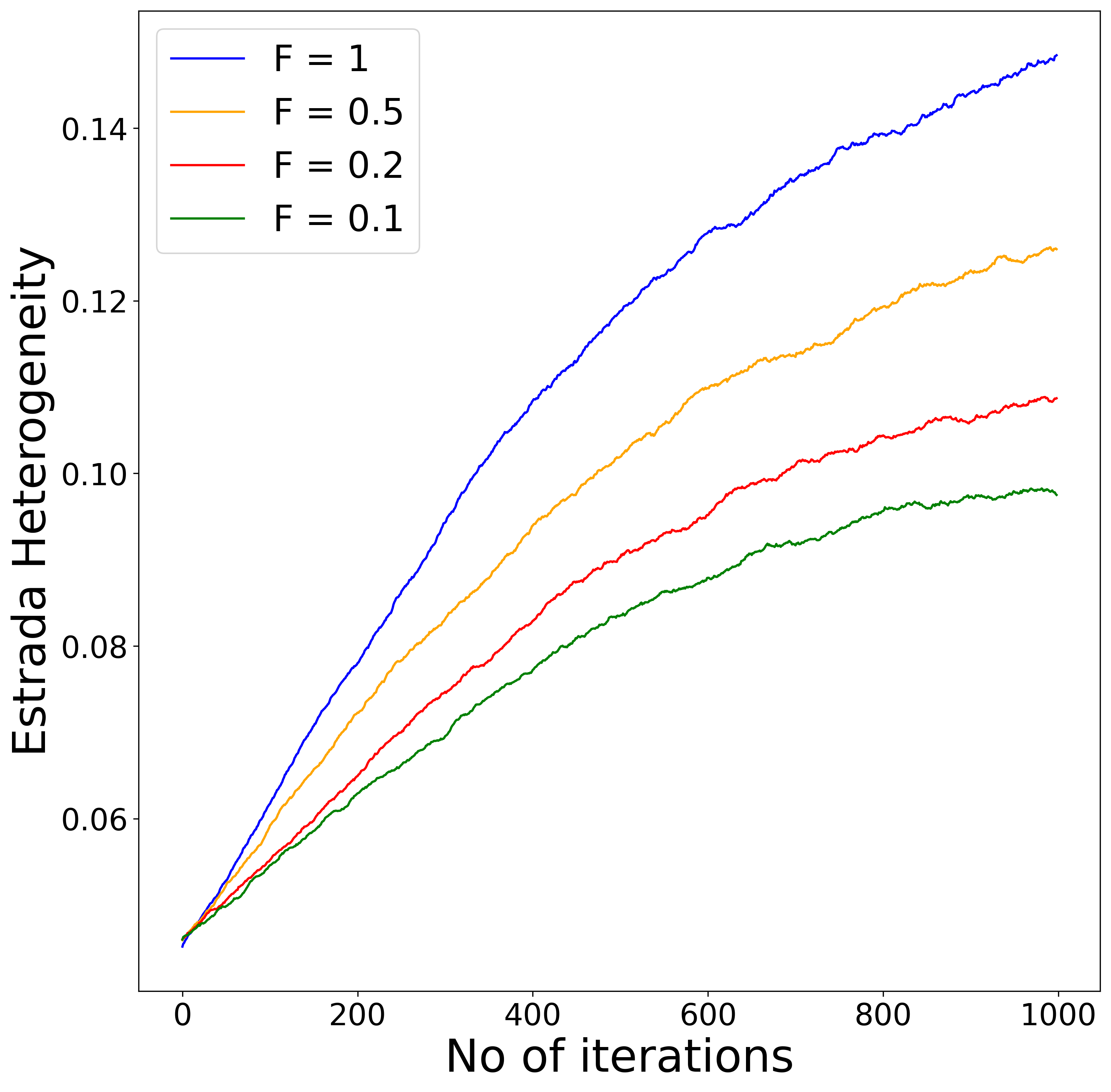}
    \caption{\centering Estrada Heterogeneity under different rewiring rates} 
    \label{rewiring_Estrada}
\end{figure} 
\FloatBarrier

\begin{figure}[htbp]
    \centering
    \includegraphics[scale=0.25]{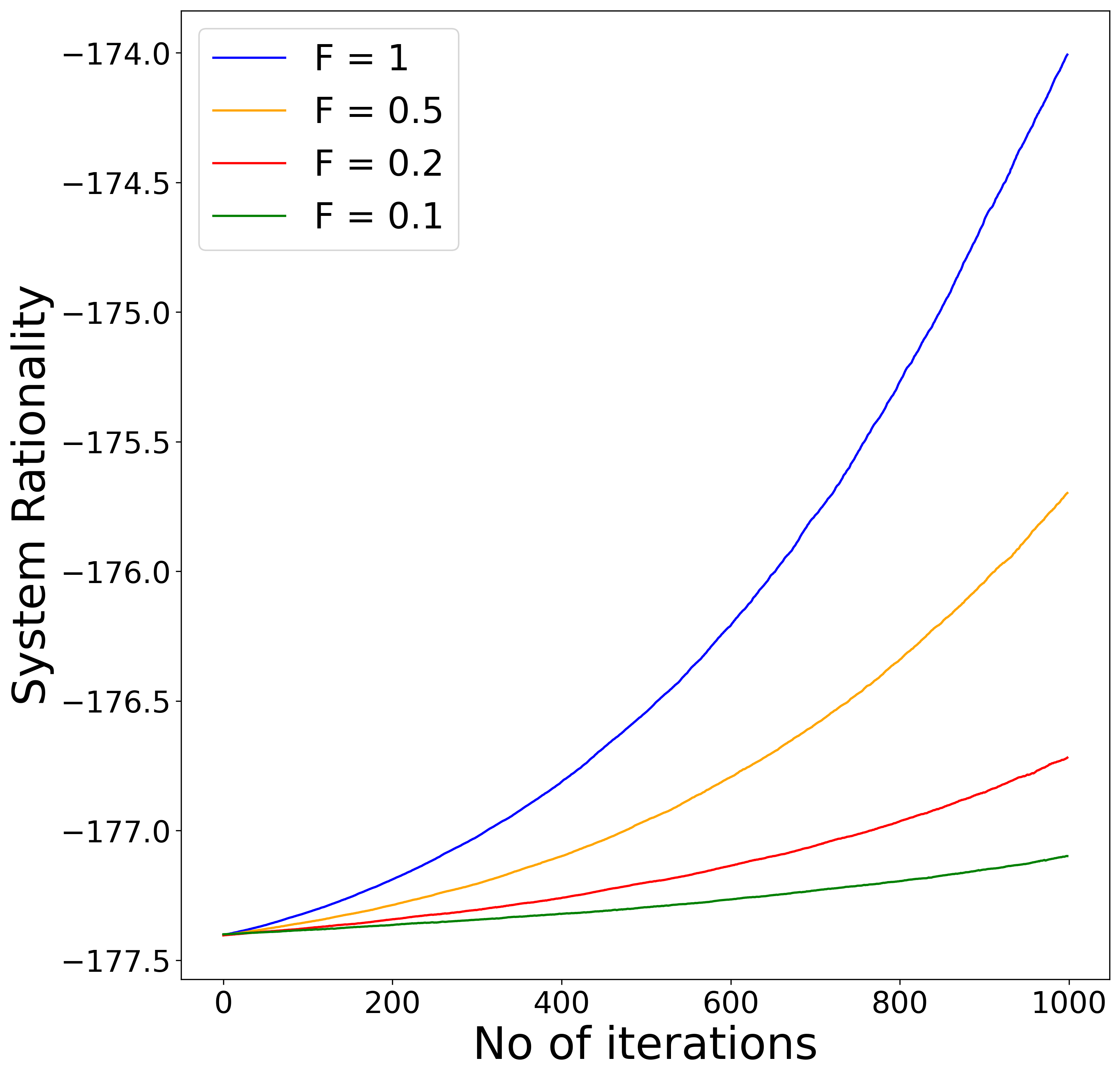}
    \caption{\centering System Rationality under different rewiring rates} 
    \label{rewiring_System_R}
\end{figure} 
\FloatBarrier

\begin{figure}[htbp]
    \centering
    \includegraphics[scale=0.25]{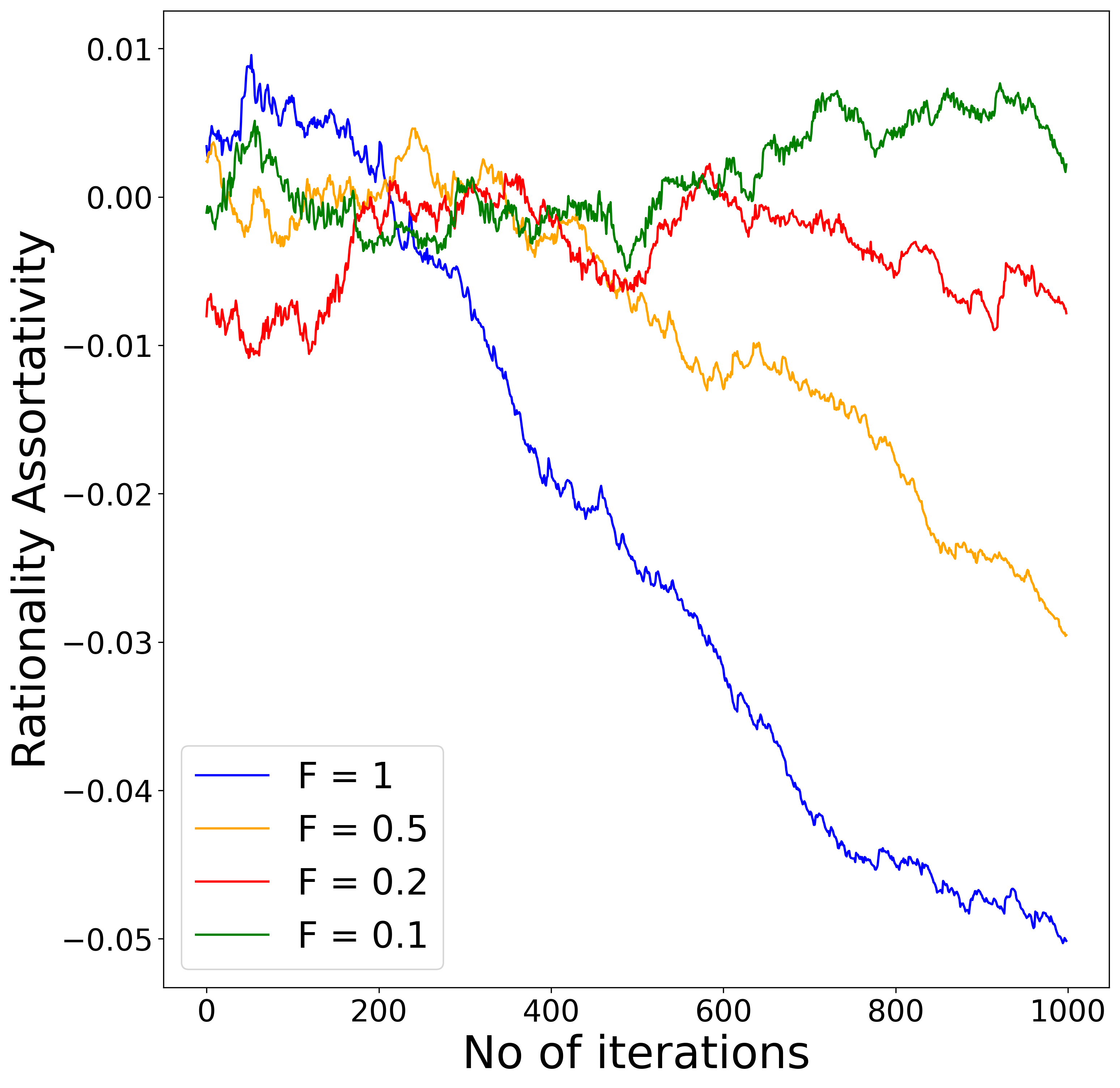}
    \caption{\centering Rationality Assortativity under different rewiring rates} 
    \label{rewiring_R_Assort}
\end{figure} 
\FloatBarrier

\begin{figure}[htbp]
    \centering
    \includegraphics[scale=0.25]{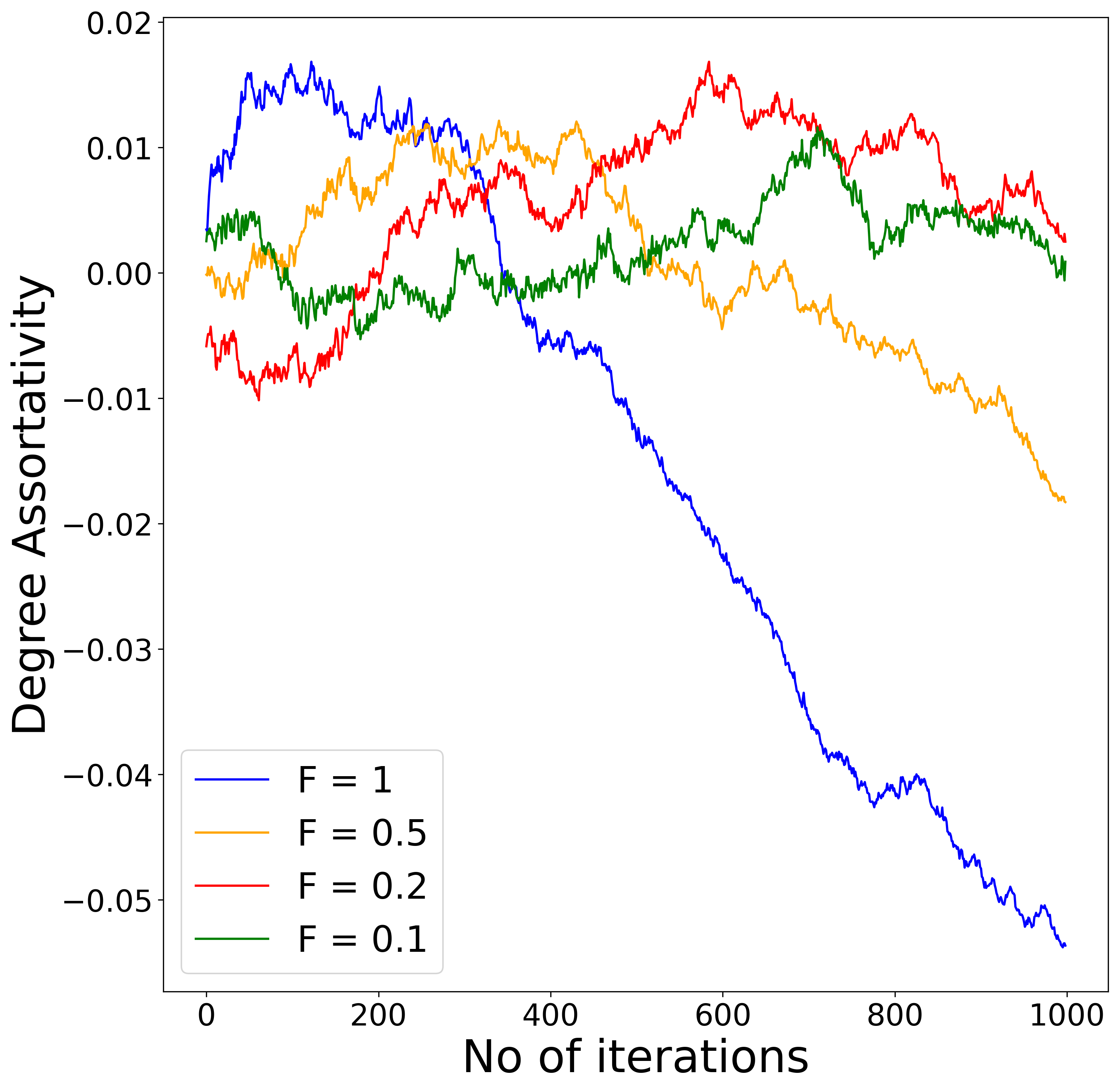}
    \caption{\centering Degree Assortativity under different rewiring rates} 
    \label{rewiring_D_Assort}
\end{figure} 
\FloatBarrier

Figures~\ref{rewiring_AvgR}--\ref{rewiring_D_Assort} demonstrate the effects of varying rewiring rates ($F$) on network evolution and system rationality.

Higher rewiring rates accelerate topological adaptation, as shown in Figures~\ref{rewiring_Powerlaw} and \ref{rewiring_Estrada}, where the network progressively evolves into a core-periphery structure. This transition is marked by a steeper power-law exponent ($\gamma \approx 1.6$--1.8) and increased Estrada heterogeneity, indicating the emergence of central hubs dominating the network’s communicability landscape.

In addition, higher rewiring rates lead to significant increase of system rationality (Figure~\ref{rewiring_System_R}), as the network quickly reorganizes to optimize interactions. However, this comes with a decline in both rationality assortativity (Figure~\ref{rewiring_R_Assort}) and degree assortativity (Figure~\ref{rewiring_D_Assort}), reflecting increasing disassortative mixing—where central, highly rational nodes connect more frequently with less rational, peripheral nodes.

These results suggest that the rewiring rate acts as a network-level learning rate, governing the speed of structural adaptation. Lower rewiring rates allow the network to evolve more gradually into classical scale-free topologies, characterized by higher assortativity and less-skewed power-law distributions. In contrast, higher rewiring rates rapidly push the network toward core-periphery structures, resulting in highly skewed power-law distributions with a few dominant hubs.

\subsection{Comparative Summary and Emergent Patterns}

Table~\ref{tab:summary} provides a comparative overview of the network structures and dynamics resulting from different learning and rewiring rate settings. The findings illustrate how individual learning rates and network-level rewiring rates jointly shape the emergent topology and systemic rationality, when the expected rationality is derived using the exponential formulation.





\FloatBarrier

\begin{table}
\centering
\caption{Comparative Summary of Learning and Rewiring Effects on Network Topology (Exponential Rationality Function)}
\label{tab:summary}
\begin{tabular}{|p{2.8cm}|p{2.1cm}|p{1.7cm}|p{2.3cm}|p{2.8cm}|}
\hline
\makecell{\textbf{Learning}\\\textbf{Configuration}} & 
\makecell[c]{$\boldsymbol{\gamma}$} & 
\makecell{Estrada\\Heterogeneity} & 
Assortativity & 
\makecell{Emergent\\Topology} \\
\hline

\makecell[l]{Low uniform\\learning rate\\($\alpha = 0.1$–$0.3$)} & 
\makecell{Moderate\\(1.4–1.5)} & 
\makecell{Medium\\(0.1–0.2)} & 
\makecell{High\\(0 to -0.10)} & 
\makecell{Moderately\\scale-free} \\
\hline

\makecell[l]{High uniform\\learning rate\\($\alpha = 0.7$-$0.9$)} & 
\makecell{High\\(1.5–1.6)} & 
\makecell{High\\(0.25–0.4)} & 
\makecell{Low\\(–0.15 to –0.40)} & 
\makecell{Core\\periphery} \\
\hline

\makecell[l]{Normal dist. LR\\($\mu = 0.1$, $\sigma = 0.1$-$0.3$)} & 
\makecell{Moderate\\(1.4–1.5)} & 
\makecell{Medium\\(0.2–0.3)} & 
\makecell{Medium\\(-0.10 to -0.20)} & 
\makecell{Mildly\\scale-free} \\
\hline

\makecell[l]{Normal dist. LR\\($\mu = 0.1$, $\sigma = 0.4$-$0.8$)} & 
\makecell{High\\(1.5-1.6)} & 
\makecell{High\\(0.3–0.4)} & 
\makecell{Low\\(–0.20 to –0.40)} & 
\makecell{Hub-dominated\\mixed} \\
\hline

\makecell[l]{Low rewiring rate\\($F = 0.1$–$0.2$)} & 
\makecell{Moderate\\(1.43–1.44)} & 
\makecell{Low\\(0.05–0.10)} & 
\makecell{High\\(+0.01 to -0.01)} & 
\makecell{Moderately\\scale-free} \\
\hline

\makecell[l]{High rewiring rate\\($F = 0.5$–$1$)} & 
\makecell{Moderate\\(1.44–1.46)} & 
\makecell{Medium\\(0.10–0.15)} & 
\makecell{High\\(–0.01 to –0.10)} & 
\makecell{Core\\periphery} \\
\hline

\end{tabular}
\end{table}

\FloatBarrier

From the above, two consistent patterns emerge:
\begin{itemize}
    \item \textbf{Node-level learning rate} (individual learnability) governs how quickly agents converge to rational behavior, which in turn influences whether centralized hubs emerge.
    \item \textbf{Network-level learning rate} (rewiring rate $F$) determines the pace of topological adaptation. High rewiring rates mimic rapid structural learning, accelerating convergence toward core-periphery structures.
\end{itemize}

\noindent
In other words, \textit{Lower learning and rewiring rates} promote moderately centralized, scale-free topologies, while \textit{higher rates} lead to strongly centralized core-periphery formations—demonstrating how both node-level cognition and network-level adaptation co-determine the emergent architecture of rational systems.

\section{Discussion}
\label{sec:discussion}

This study presents a unified simulation framework that models the co-evolution of individual behavior and network structure through the dual lens of learning rate heterogeneity. Specifically, we introduce two dimensions of learning: \textbf{node-level learning rates}, which govern how quickly individual agents adapt their rationality, and \textbf{network-level learning rates}, operationalized as the rewiring rate, which determine how frequently the network adapts its topology based on system-level feedback. By jointly modeling these two learning dynamics, we provide new insights into the mechanisms that drive the emergence of diverse network topologies in socio-economic systems.

A key contribution of this work is the reconciliation of seemingly contrasting results in the literature. Prior research by Kasthurirathna and Piraveenan~\cite{kasthurirathna2015emergence} has shown that adaptive socio-economic networks tend to evolve into \textit{scale-free topologies} as system rationality improves. These structures, characterized by a power-law degree distribution and long-tailed connectivity~\cite{barabasi1999emergence}, enable robust and efficient information flow. On the other hand, Roman and Brede~\cite{roman2017topology} has argued that adaptive networks can converge toward more \textit{core-periphery structures}, especially when rewiring is driven by rational performance-based optimization. Our framework demonstrates that both patterns can emerge within a single theoretical model, depending on the configuration of learning rates.

When \textbf{node-level learning rates} are lower and relatively homogeneous, networks tend to evolve toward classical scale-free structures. However, introducing heterogeneity in these learning rates—particularly when a small subset of nodes learns rapidly—drives the formation of strongly centralized, hub-dominated structures. Similarly, increasing the \textbf{network-level learning rate} (i.e., the rewiring rate) accelerates structural adaptation, leading to pronounced core-periphery topologies~\cite{borgatti2000models}. These dual learning parameters thus act as dials: by tuning the speed and variability of adaptation, the system transitions between distributed and centralized network forms.

The resulting topologies closely resemble many real-world socio-economic networks. For example, \textit{global airline networks}~\cite{guimera2005worldwide}, \textit{financial interbank lending networks}~\cite{iori2008network}, and \textit{urban transportation systems}~\cite{von2009surprising} often display both \textit{scale-free properties} and \textit{core-periphery organization}. In these systems, a few key hubs (e.g., major cities or financial institutions) dominate the connectivity landscape, while peripheral nodes maintain sparse links, creating hybrid structures that are both efficient and vulnerable. Our findings suggest that such hybrid structures may be a natural consequence of heterogeneous learning capacities among agents and differing rates of network adaptation.

While the simulation framework offers novel insights, several limitations warrant consideration. First, the model assumes that agents play a single repeated game (Prisoner's Dilemma) using QRE-based strategy selection~\cite{mckelvey1995quantal}. This restricts the behavioral richness of agents, as real-world decision-making often involves evolving strategies, bounded memory, or learning from peers~\cite{macy2002learning}. Second, learning rates are either randomly sampled or tied to node degree, omitting potentially important drivers such as agent experience, resource availability, or social influence~\cite{rogers2003diffusion}. Third, the network rewiring process assumes global visibility of system rationality to guide structural change—an assumption that may not hold in decentralized systems. Additionally, the simulations are conducted on a single-layer static graph, whereas many socio-economic systems are inherently multi-layered (e.g., combining social, informational, and financial interactions) and subject to exogenous shocks~\cite{kivela2014multilayer}.

Future work can expand the scope of this model in several directions. Other centrality measures, such as eigenvector centrality or betweenness centrality, can also be employed to adjust the learning rate, offering alternative perspectives on how an agent’s structural influence within the network may shape its rationality. Further, incorporating alternative learning mechanisms such as reinforcement learning, Bayesian updating, or imitation dynamics~\cite{traulsen2007stochastic} could yield a more realistic depiction of agent adaptation. Second, exploring \textit{temporal or multi-layer networks} would allow us to investigate how learning rates influence interdependent systems with different types of interactions. Third, empirical calibration of the model using data from real-world networks—such as digital platforms, supply chains, or public health systems—could help validate the role of learning heterogeneity in shaping structure. Finally, policy simulations could be conducted to evaluate how targeted interventions (e.g., slowing down or accelerating the learning of certain nodes, altering rewiring thresholds) might influence overall system efficiency, equity, or resilience.

In summary, this work advances the theoretical understanding of adaptive socio-economic networks by introducing a dual-learning perspective. By explicitly modeling both individual learnability (through node-level learning rates) and collective structural learning (through network-level rewiring rates), we provide a framework that not only captures known dynamics like scale-free emergence and hub dominance, but also explains how combinations of learning heterogeneity can generate hybrid topologies observed in real systems. This opens new pathways for analyzing and designing adaptable and intelligent socio-economic infrastructures.

\bibliography{references}
\end{document}